\documentclass[fleqn,usenatbib,onecolumn]{mnras}
\usepackage{newtxtext,newtxmath}
\usepackage[T1]{fontenc}
\usepackage{ae,aecompl}

\usepackage{graphicx}	% Including figure files
\usepackage{amsmath}	% Advanced maths commands
\usepackage{pdflscape}  % For landscape table
\title[White Dwarfs in SDSS DR16]{White dwarf and subdwarf stars in the Sloan Digital Sky Survey Data Release 16}
   \author[Kepler et al.]{S. O. Kepler$^{1}$\thanks{kepler@if.ufrgs.br},
                Detlev Koester$^{2}$,
                Ingrid Pelisoli$^{3}$,
                Alejandra D. Romero$^{1}$,
                Gustavo Ourique$^{1}$\\
$^{1}$Instituto de F\'{\i}sica, Universidade Federal do Rio Grande do Sul,
              91501-900  Porto-Alegre, RS, Brazil\\
$^{2}$Institut f\"ur Theoretische Physik und Astrophysik, Universit\"at Kiel, 24098 Kiel, Germany\\
$^{3}$Department of Physics, University of Warwick, Gibbet Hill Road, Coventry, CV4 7AL, UK\\
}

\date{Accepted Received 2021 June 15}

\pubyear{2021}

\begin{document}
\label{firstpage}
\pagerange{\pageref{firstpage}--\pageref{lastpage}}
\maketitle

\begin{abstract}
White dwarfs are the end state of the evolution of more than 97\% of all stars, and therefore carry information on the structure and evolution of the Galaxy through their luminosity function and initial-to-final mass relation. 
Examining the new spectra of all white or blue stars in the Sloan Digital Sky Survey Data Release 16, we report the spectral classification of 2410 stars, down to our identification cut-off of signal-to-noise ratio equal to three. We newly identify 
1404 DAs,
189 DZs,
103 DCs, 12 DBs, and 9 CVs. The remaining objects are a mix of carbon or L stars (dC/L), narrow-lined hydrogen-dominated stars (sdA), dwarf F stars and P Cyg objects.
As white dwarf stars were not targeted by SDSS DR16, the number of new discoveries is much smaller than in previous releases. We also report atmospheric parameters and masses for a subset consisting of 555 new DAs, 10 new DBs, and 85 DZs for spectra with signal-to-noise ratio larger than 10.
\end{abstract}


%%%%%%%%%%%%%%%%%%%%%%%%%%%%%%%%%%%%%%%%%%%%%%%%%%

%%%%%%%%%%%%%%%%% BODY OF PAPER %%%%%%%%%%%%%%%%%%

\section{Introduction}

White dwarf stars are the final stage of evolution for all stars formed with initial masses below around
7--11~$M_{\odot}$, depending on metallicity \citep[e.g][]{Ibeling13,Doherty14,Woosley15,Williams18,Ramos18},
which represent more than 97\% of all stars in our Galaxy. White dwarf stars have masses below the Chandrasekhar limit, around $1.4~M_\odot$
\citep[e.g.][]{Chandra31,Chandra64,kilic21},
%Caiazzo21
and their mean mass is around $0.6~M_\odot$ \citep[e.g.][]{Koester79,tremblay20}.
They are also possible outcomes of the evolution of multiple systems, with 25--30 per cent of white dwarfs 
estimated to be the result of
mergers \citep[e.g.][]{Toonen17}. 
White dwarfs with masses lower than 0.3--0.45~$M_{\sun}$ are generally explained as the result of close binary evolution \citep[e.g.][]{Marsh95,Kilic2007}, because single progenitors of such low-mass white dwarfs have main sequence lifetimes exceeding the age of the Universe. The formation
mechanism of the so-called extremely-low mass white dwarfs (ELMs) -- those with masses below $\simeq 0.2-0.3 M_\odot$ \citep[e.g.][and references therein]{Sun18, Calcaferro18} -- is similar to that proposed to explain  hot subdwarf stars \citep[e.g.][]{Heber16}:
the outer envelope is lost after a common envelope or a stable Roche-lobe overflow phase, leaving the stellar core exposed \citep[e.g.][]{li2018}. 

White dwarfs do not present ongoing core nuclear burning,
even though residual shell burning may occur depending on the thickness of their outer hydrogen layer.
ELM models suggest that they present  residual burning before reaching the final white dwarf cooling track \citep{Corsico12,Istrate16}. This happens in the pre-ELM phase \citep{Maxted14a, Maxted14b},
which can cause them to brighten to luminosities comparable to main sequence and even horizontal branch stars \citep[e.g.][]{RRLyra026}.

Around 80\% of all white dwarfs show solely hydrogen lines, and are classified as spectral class DA.
This occurs because the timescales for gravitational settling are of the order
of a few million years or smaller \citep[][]{Schatzman,Michaud}, leading to a generally simple atmospheric composition,
with the lightest element available on the surface, except for effective temperatures above $\sim$50\,000~K.
The spectral class of the majority remaining is DB, if only He\,I lines are present, and DO if He\,II lines are visible
--- typically with $T_\mathrm{eff} \gtrsim 40\,000$~K. Very cool white dwarfs  --- $T_\mathrm{eff} \lesssim 5\,000$~K for H atmosphere, 
$T_\mathrm{eff} \lesssim 11\,000$~K for He atmosphere --- show featureless spectra
and are classified as DCs. 
A substantial fraction \citep[20--50 per cent,][]{Zuckerman03,Koester14,Hollands17,Hollands18} of white dwarfs show contamination by metals, which can only be explained by ongoing accretion,
except for very hot objects ($T_\mathrm{eff} \gtrsim 50\,000$~K), where radiative levitation can still play a significant role \citep[e.g][]{Bruhweiler83,Chayer89,Barstow14};
a Z is added to the spectral classification to flag metal pollution. In rare cases, for stars classified as DQs, carbon may be dragged to the surface by convection 
\citep[e.g.][]{Koester82,Pelletier86,Blouin19}. Cool DQs show spectra similar to dwarf carbon (dC) stars, which are themselves believed to be one outcome of binary evolution
\citep[e.g.][]{Whitehouse18}. Even rarer are those white dwarfs with spectra dominated by oxygen lines, classified as DS \citep{kurtis19}.

In this paper we extend the work of \citet{dr7} and \citet{dr10,dr12,dr14}, continuing the search for new spectroscopically confirmed white dwarf and subdwarf stars in the 
data release 16 of the Sloan Digital Sky Survey (SDSS) \citep[SDSS DR16,][]{Ahumada20}, which contains SDSS observations through August 2018. 
Spectroscopy allows estimations of $T_\mathrm{eff}$, $\log~g$, and abundances, serving
as a valuable resource for studying stellar formation and evolution in the Milky Way \citep[e.g][]{winget1987, BSL, Liebert2005, Tremblay14}. 
As a by-product, we also identify cataclysmic variables (CVs) --- white dwarfs with ongoing mass exchange from a companion, and presenting emission lines, generally of hydrogen and helium --- and dC stars, due to their spectral similarity with carbon-rich white dwarfs. These dC stars \citep{Roulston18}, as well as hot subdwarfs and ELMs,
hold potential to shed light on the poorly understood process of close binary evolution.

\section{Data analysis}

\subsection{Identification of the candidates}
\begin{figure}
   \centering
   \includegraphics[width=\linewidth]{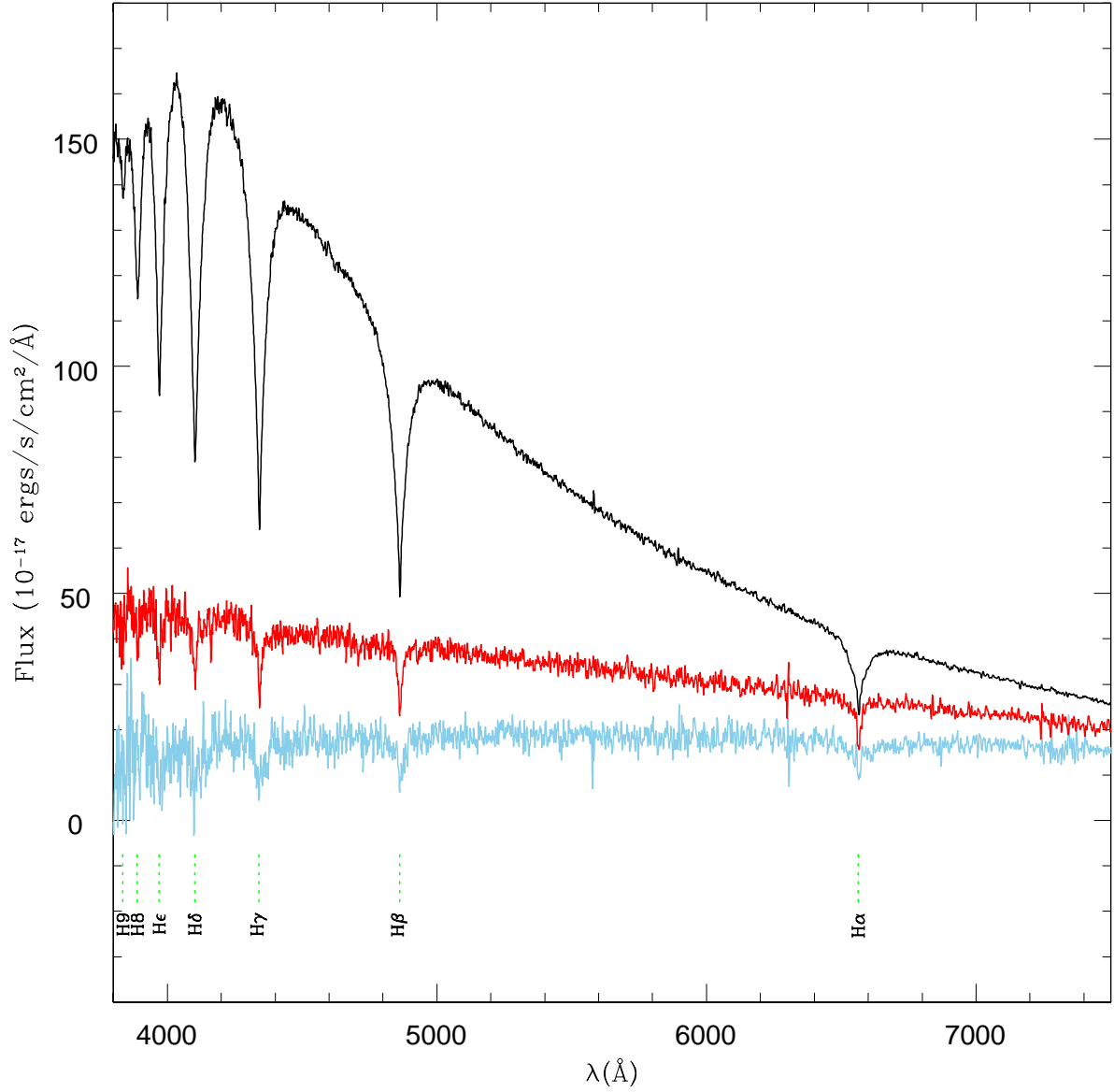}
      \caption{Spectra of three DAs, from top to bottom, S/N$_g$=70, SDSSJ075144.06+223004.80, P-M-F 11112-58428-0632, g=16.639; S/N$_g$=13, the mean S/N,
      SDSS~J142707.81+381640.81, P-M-F 10752-58488-0162, g=18.919, scaled for display,
      and S/N$_g$=3, SDSS~J132411.33+340028.76, P-M-F 10254-58514-0076, g=18.319, scaled, to show the range of S/N we classified.}
   \label{dasn}
\end{figure}

This paper follows on the search for new spectroscopically identified white dwarf stars from spectra obtained by the Sloan Digital Sky Survey \citep{dr4,dr7,dr10,dr12,dr14}.
We started with all the optical spectra obtained after the SDSS Data Release 14.
We selected all newly observed spectra within the colour selection of \citet{dr7}, $\approx 78\,000$, and all spectra classified by the SDSS spectral pipeline as WHITE\_DWARF, A, B, OB or O stars, or
CV (cataclysmic variables), 49\,667 spectra. There is partial overlap between the two samples.
In addition, we performed an automated algorithm search for spectra similar to previously selected DA and DB training samples, as described in \citet{dr10,dr12,dr14},
on all the $\approx 1\,970\,000$ new optical spectra from DR16, which include the sample of already selected spectra. This resulted in the selection of 858  further spectra that were not included in the colour or pipeline classification samples.  As white dwarfs were not specifically targeted by DR16, the number of new white dwarfs is smaller than in the previous data releases.
Due to the overlaps, we examined these $\approx 128\,000$ selected spectra by eye,
to identify broad line spectra characteristic of
white dwarfs, hot subdwarfs, and dCs. Using an identification cut-off of (S/N)$_g\geq 3$, where 
(S/N)$_g$ is the signal-to-noise parameter in the g-band from the SDSS spectra reduction
pipeline, we identified 2410 
spectra containing white dwarf, subdwarf, CV and dC stars. In Fig.~\ref{dasn} we display three spectra with (S/N)$_g$=70, 13, and 3, for comparison of the range of S/N we classified.
The redder dC/L stars, and some DZs, do not reach (S/N)$_g\geq 3$, but show significant flux at redder wavelengths.
As in previous Data Releases, our visual inspection showed that most objects
with SDSS spectra, proper motion smaller than 30~mas/yr,
and magnitude $g>20$ are in fact galaxies, from their composite spectrum, high red-shifted lines, or broad emission lines.

\subsection{Spectral Classification}

DR16 uses improved flux-calibration, with atmospheric differential refraction corrected on a per-exposure basis following the recipe described in \citet{Margala16},
and improved co-addition of individual exposures.
The Stellar Parameters Pipeline, which we used in our initial spectral class selection, are from \citet{lee08a, lee08b} and  \citet{AllendePrieto08}.

The wavelength coverage	is from 3650 to 10\,400~\AA\ for the BOSS spectrograph, with a
resolution of 1500 at 3800~\AA\ and 2500 at 9000~\AA, and a
wavelength calibration better than 5 km/s.
All the spectra used in our analysis
were processed with the spectroscopic reduction pipeline version
v5\_13\_0.
These
RUN2D numbers denote the version of extraction and redshift-finding code used.
In all SDSS spectral line descriptions, vacuum wavelengths are used.
The wavelengths are shifted such that measured velocities are relative to the solar system barycentre at the mid-point of each 15-minute exposure. 

We were conservative in labelling a spectrum as a clean
DA or DB, considering we are interested in obtaining accurate mass distributions for
our DA and DB stars.
we therefore add additional subtypes
and uncertainty notations (:) if we saw signs of other elements, unresolved companions, or
magnetic fields (H) in the spectra. 
While some of our mixed white dwarf
subtypes would possibly be identified as clean DAs or DBs with better
signal-to-noise spectra, few of our identified clean DAs or DBs would
likely be found to have additional spectral features within our detection
limit.

We looked for the following features to aid in the
classification for each specified white dwarf subtype:
\begin{itemize}
\item Balmer lines --- normally broad and with a Balmer decrement
[DA but also DAB, DBA, DZA, and subdwarfs]
\item HeI $4\,471$\AA\ [DB, sdB]
\item HeII $4\,686$\AA\ [DO, PG1159, sdO]
\item C2 Swan band or atomic CI lines [DQ]
\item CII $4\,367$\AA\ [HotDQ]
\item CaII H \& K  [DZ, DAZ, DBZ]
\item Zeeman splitting [magnetic white dwarfs]
\item featureless spectrum with significant proper motion [DC]
\item flux increasing in the red [binary, most probably M companion]
\item OI $6\,158, 7\,774, 8\,448$\AA\ [DS, oxygen dominated]
\item H and He emission lines [CVs and M dwarf companions]
\end{itemize}
\goodbreak

\begin{table}
	\centering
\caption{Classification of 2\,410 spectra in Table~\ref{tab:all.1}, including 30 known CVs and one new CV with two spectra.
	\label{tab:all}}
	\begin{tabular}{rl} 
		\hline
Number & Type \\
		\hline
   1404 & DA \cr
    12 & DB \cr
    103 & DC \cr
    189 & DZ \cr
     41 & CV \cr %only 9 new
%      2   & DS \\
      & \cr
     301 & dC/L \cr %dwarfC
   320 & sdA \cr
   16 & F\cr
   4 & P Cyg\cr
   1 & binary\cr
		\hline
	\end{tabular}
\end{table}

\begin{table}
%\begin{minipage}{\linewidth}
%\begin{landscape}
\caption{Spectral classification for 2\,410 white dwarfs and subdwarfs in SDSS DR16. The complete table is available electronically (dr16tabedr3.csv).}
\label{tab:all.1}
\tiny
\setlength{\tabcolsep}{2pt}
\centering
\begin{tabular}{lcccccccccccccccccccrrrrl}
SDSS J&Plate-MJD-Fiber&S/N$_g$&u&$\sigma_u$&g&$\sigma_g$&r&$\sigma_r$&i&$\sigma_i$&z&$\sigma_z$&E(B-V)&ppm&$\ell$&b&DR2par&$\sigma_\pi$&EDR3par&$\sigma_\pi$&ppm$_G$&G&BP-RP&Type\cr
&&&(mag)&(mag)&(mag)&(mag)&(mag)&(mag)&(mag)&(mag)&(mag)&(mag)&(mag)&(mas/yr)&(deg)&(deg)&(mas)&(mas)&(mas)&(mas)&(mas/yr)&(mag)&(mag)&\cr
000007.28+074944.83&11279-58449-0954&06&20.938&0.095&20.481&0.025&20.743&0.036&20.911&0.064&21.306&0.325&0.043&000.0&101.5&-52.9&2.663&1.168&2.866&0.989&30&20.627&0.384&DA\cr
000035.47+084315.04&11277-58450-0316&06&21.335&0.151&21.422&0.051&21.208&0.059&21.466&0.132&20.497&0.239&0.051&005.4&102.2&-52.1&0.547&0.822&0.339&0.624&02&20.150&0.526&DA\cr
000035.59-001115.94&09403-58018-0465&07&23.039&0.347&20.271&0.031&18.813&0.017&18.334&0.026&18.118&0.039&0.030&065.8&096.5&-60.4&0.644&0.378&1.403&0.383&64&18.834&1.753&dC\cr
000155.08+101323.36&11277-58450-0764&05&21.077&0.072&20.744&0.030&20.907&0.034&21.046&0.060&21.132&0.216&0.064&000.0&103.4&-50.8&&&&&&20.775&0.250&DA\cr
000211.15+092909.09&11277-58450-0822&07&21.239&0.079&20.651&0.040&20.434&0.029&20.025&0.035&19.354&0.046&0.098&028.5&103.2&-51.5&3.267&0.778&2.404&0.650&27&20.196&1.175&DA+M\cr
000211.25+114243.06&11561-58485-0546&12&19.825&0.049&19.436&0.033&19.542&0.022&19.745&0.029&19.909&0.092&0.063&045.1&104.2&-49.4&2.829&0.431&3.179&0.325&36&19.459&-0.007&DA\cr
000212.38+110905.43&11561-58485-0513&03&22.121&0.161&21.300&0.039&20.877&0.039&20.741&0.052&20.773&0.171&0.057&000.0&104.0&-49.9&&&&&&20.904&0.079&sdA\cr
000219.56+085915.02&11277-58450-0872&12&21.439&0.096&19.869&0.028&19.116&0.015&18.869&0.018&18.737&0.032&0.058&002.9&103.0&-52.0&0.000&0.436&0.025&0.292&1&19.150&1.027&DZ\cr
000221.17+090551.44&11277-58450-0830&03&22.153&0.159&21.540&0.053&20.998&0.035&20.943&0.047&20.944&0.177&0.058&000.0&103.1&-51.9&&&&&&21.064&0.921&sdA\cr
000329.78+092751.00&11277-58450-0882&05&21.874&0.133&21.064&0.039&20.592&0.033&20.370&0.044&20.588&0.144&0.099&033.4&103.7&-51.6&0.712&0.970&1.583&0.795&3&20.537&0.835&DZ:\cr
000348.97+100936.80&11561-58485-0419&04&22.807&0.484&20.933&0.039&21.002&0.057&21.028&0.072&21.313&0.406&0.089&000.0&104.1&-51.0&&&&&&20.885&0.455&DA\cr
000356.67+110603.64&11561-58485-0612&15&19.651&0.034&19.448&0.025&19.643&0.021&19.880&0.030&20.069&0.107&0.057&034.7&104.6&-50.1&4.739&0.507&5.288&0.366&37&19.519&0.018&DA\cr
000426.89+104059.93&11561-58485-0396&04&21.611&0.099&21.015&0.036&21.095&0.042&21.071&0.060&20.948&0.194&0.093&000.0&104.6&-50.5&&&&&&21.041&-0.525&DA\cr
000444.53+085938.62&11277-58450-0939&06&21.219&0.078&20.670&0.034&20.787&0.033&21.035&0.061&21.233&0.261&0.073&000.0&103.9&-52.1&1.722&1.676&0.707&1.572&25&20.697&0.212&DA\cr
000457.14+010937.47&09403-58018-0843&13&22.101&0.180&19.564&0.019&18.396&0.014&18.005&0.019&17.765&0.034&0.025&043.9&099.5&-59.6&0.000&0.379&0.297&0.230&42&18.436&01.507&dC\cr
000710.54+114549.41&11561-58485-0790&03&23.080&0.476&21.426&0.051&20.852&0.044&20.776&0.056&20.521&0.169&0.084&000.0&106.1&-49.7&&&&&&20.855&0.357&sdA\cr
001012.54+105816.47&11561-58485-0099&02&22.246&0.332&21.953&0.138&21.569&0.140&21.443&0.142&21.491&0.352&0.087&000.0&106.9&-50.6&1.997&1.306&1.533&0.951&1&20.594&0.794&DA:\cr
001140.11+103759.11&11561-58485-0026&08&21.050&0.077&20.288&0.022&19.936&0.022&19.868&0.030&19.745&0.064&0.061&009.0&107.3&-51.0&0.720&0.813&0.236&0.632&2&19.969&0.388&sdA\cr
001202.04+064945.88&11309-58428-0015&19&20.157&0.040&19.034&0.021&19.070&0.016&19.144&0.020&19.136&0.041&0.057&001.2&105.8&-54.7&0.846&0.645&0.907&0.404&4&18.996&0.329&sdA\cr
001245.81-010522.66&09402-58039-0891&03&24.122&0.706&21.191&0.040&19.558&0.016&19.069&0.024&18.886&0.037&0.033&050.3&101.7&-62.4&0.826&0.570&0.357&0.417&49&19.650&1.798&dC\cr
\end{tabular}
%\end{landscape}
%\end{minipage}
\end{table}

\begin{figure}
   \centering
   \includegraphics[width=\linewidth]{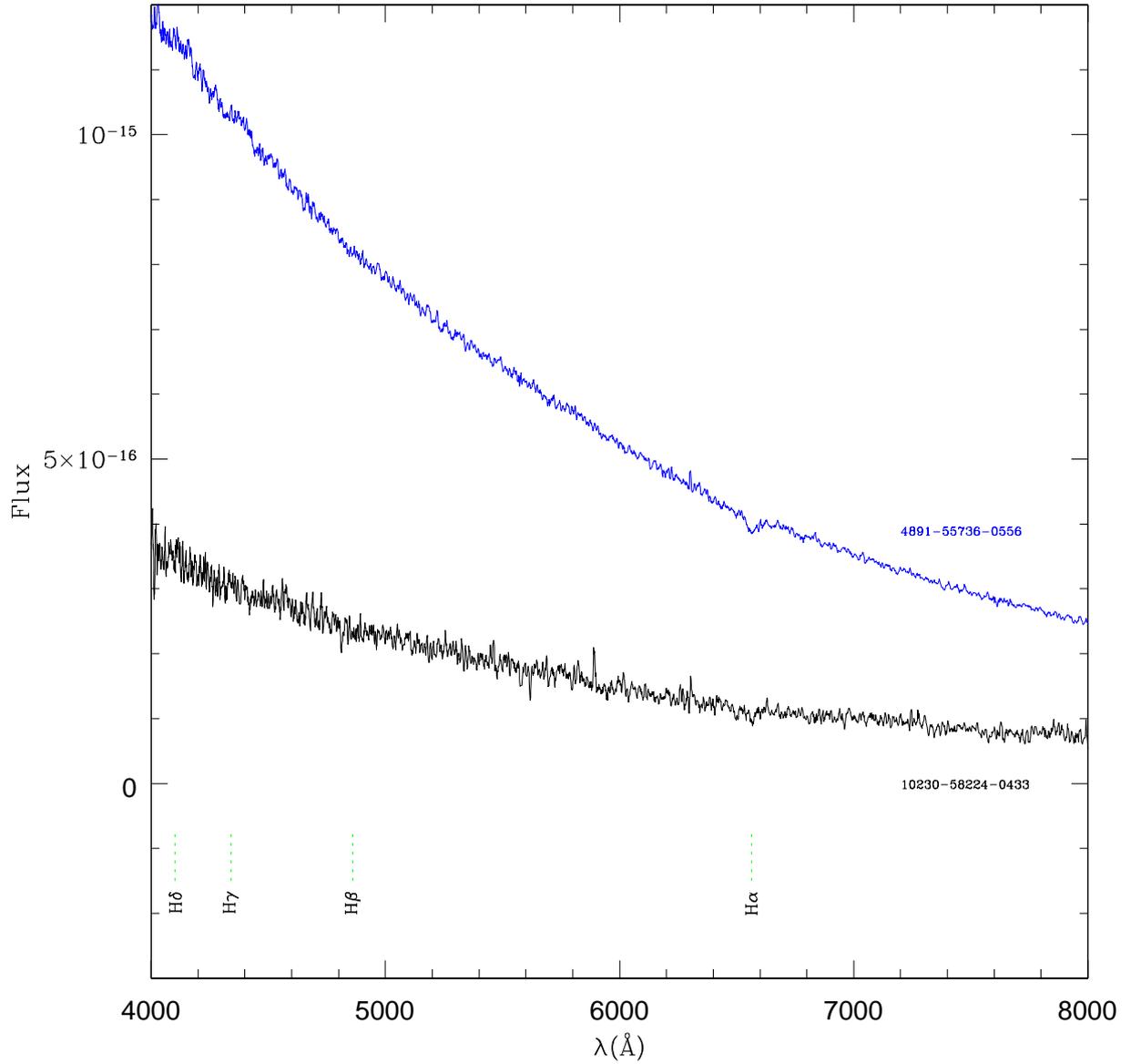}
      \caption{Spectrum of SDSS~J095018.83+340743.17, with P-M-F 10230-58224-0433 from DR16 and SDSS~J152958.12+130454.80, with P-M-F 4891-55736-0556 from DR14 in blue for comparison, two stars we classified as DA(He) because only a week H$\alpha$ is detected.}
   \label{dahe}
\end{figure}

Table~\ref{tab:all} is a tally of the 2\,410 objects we classified in Table~\ref{tab:all.1}. 1404 
objects were classified by us as new DAs and only 12 as new DBs.
Among the 1\,404 DAs, we found 8 magnetic DAs (DAH),  154 showing composite spectra with main-sequence M dwarf companions (DA+M), 46
DAZs with Ca and/or Mg lines, and 6 DABs with H and He\,I lines.
We also found one star with an extremely steep Balmer decrement, i.e. with only a broad H$\alpha$ line while the other lines are absent.
It could not be fit with a pure hydrogen grid (see section~\ref{models} below), or indicated extremely high gravity.
We find that this object is best explained as helium-rich DA, and therefore with an extremely thin H layer mixed with the underlying He, and denote it DA(He), as in \citet{dr10,dr12,dr14} (see Fig.~\ref{dahe}).
\begin{figure}
   \centering
   \includegraphics[width=\linewidth]{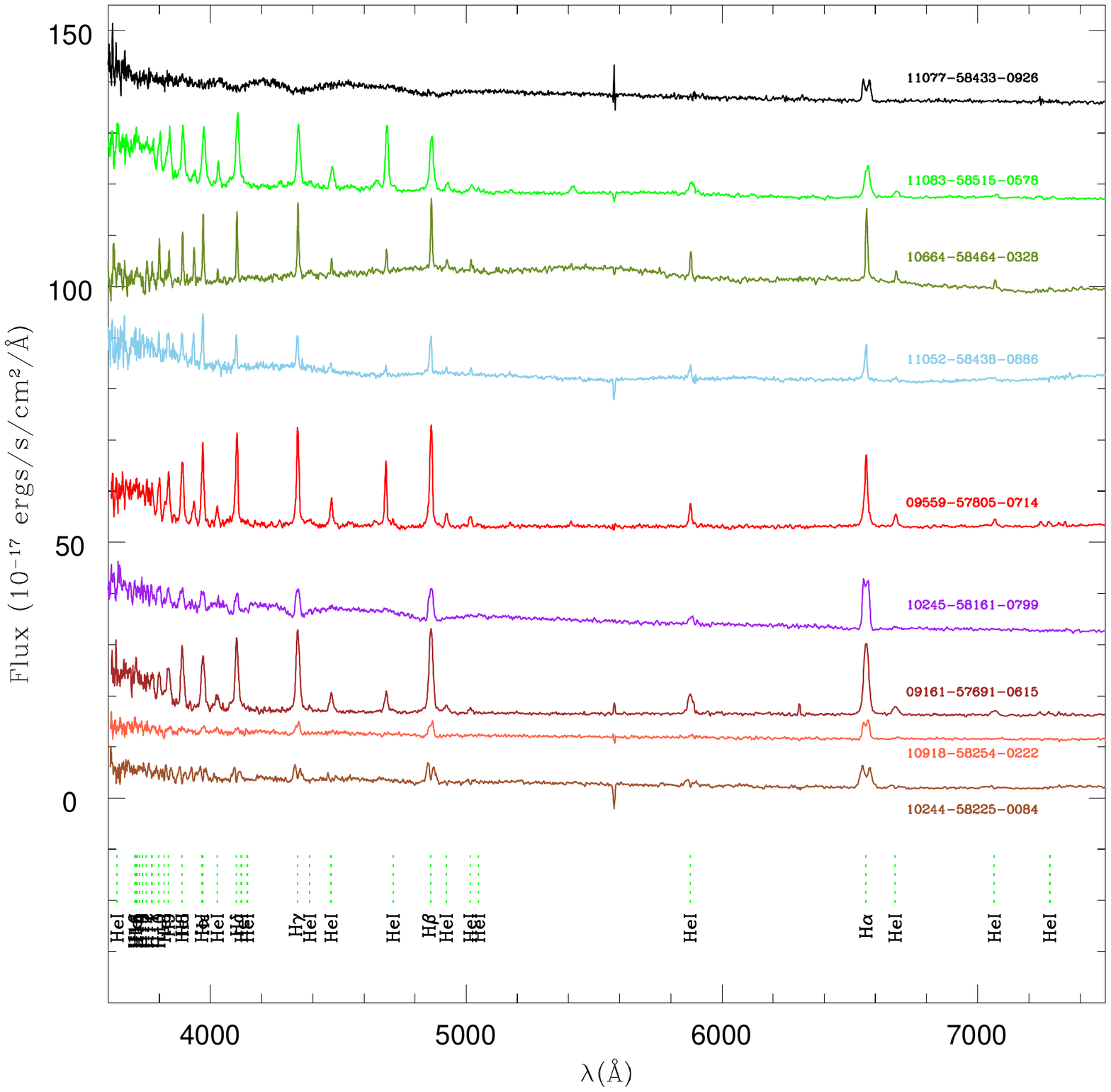}
      \caption{Spectra of the 9 new CVs we classified:
      SDSS~J012212.21+075546.84 P-M-F=11077-58433-0926,
      SDSS~J014732.86+144443.37 P-M-F=11052-58438-0886,
      SDSS~J073605.07+182709.83 P-M-F=11083-58515-0578,
SDSS~J083404.25+185416.87 P-M-F=09559-57805-0714,
SDSS~J083549.86+292636.94 P-M-F=10664-58464-0328,
SDSS~J093130.75+335651.30 P-M-F=10244-58225-0084,
SDSS~J121015.61+351334.49 P-M-F=10245-58161-0799,
SDSS~J161853.17+303845.50 P-M-F=10918-58254-0222,
SDSS~J211652.28+014144.34 P-M-F=09161-57691-0615.
      }
   \label{cv}
\end{figure}

We classified 301 spectra as dC - dwarf carbon stars, in line with \citet{Green13} and \citet{Farihi18}, but they could also be late type L stars.
We do not have spectral models for dCs or L stars, so we do not determine their properties.
All 40 CV spectra show both H and He lines, and 11 show also evidence of a disk. Of these CVs, 30 objects already have previous published spectra, and 9 are new cataclysmic variables (see Fig.~\ref{cv}). One new CV has two spectra. We kept the known CVs in the table because their spectra change with time. None shows only He lines, as expected from AM~CVns.

We classified 320 stars as sdAs, stars with spectra dominated by narrow hydrogen lines, following \citet{dr12}. 
Solar metallicity main sequence A type stars have absolute magnitudes $M_g\simeq 0$ -- 2. As stars brighter than g=14.5
saturate in SDSS, only main sequence A type stars with distance moduli larger than 12.5 are observed in SDSS, i.e., farther than 3.5~kpc.   
Because SDSS observed mainly perpendicular to the disk, i.e. galactic latitude in general larger than $30\deg$, these would be located in the halo, where most A type dwarf stars should already have evolved off the main sequence.
Most of these sdA stars are likely very low metallicity main sequence stars ($\mathrm{[Fe/H]} \lesssim -1.0$), whose spectra are dominated by hydrogen because they lack significant metals \citep[e.g.][]{chandra21}.
As their absolute magnitude, according to Gaia parallaxes, cover $M_\mathrm{G} \geq 9.5$ %(see Fig.~\ref{Fig:sda}),
they cannot be classified as normal main
sequence A type stars. Instead, they likely have masses smaller than the Sun, given their relatively low effective temperatures and location in the {\it Gaia} colour-magnitude diagram. We also note that they are hotter than sdF stars \citep{scholz15}. Some of these sdAs may be stars that lost mass due to binary interaction, resulting most probably in He core stars, precursors of ELMs \citep{Pelisoli18a,Pelisoli18b,Pelisoli18c,Pelisoli19}. %(see Section~\ref{section:sub}).

\subsection{Theoretical models and fitting methods\label{models}}

The observed spectra for pure DAs, DBs, and DZs were analysed using
theoretical LTE (local thermodynamic equilibrium) atmospheric
models. The basic principles are described in \citet{Koester10}, but many
improvements to the algorithms and models have been included since, as
described e.g. in \citet{Koester19,Koester20}. The
pure hydrogen DA models use the mixing length approximation for
convection with the parameters MLT2/$\alpha$=0.7; the grid covers the
range 5000~K $\leq T_\mathrm{eff} \leq$ 100\,000~K, 5.0 $\leq \log g \leq$ 9.5 dex (cgs).  The DB grid uses
MLT2/$\alpha=1.25$, and covers 12\,000~K $\leq T_\mathrm{eff} \leq$ 45\,000~K, 7.0 $\leq \log g \leq$9.5. 
This DB grid includes trace hydrogen with logarithmic abundances
of [H/He] = -5.0, which gives a better agreement with the H-rich atmosphere mass distribution \citep{Bergeron19,McCleery20}.
The DZ grid uses the 13 most important metals with Z<30, where the
metals are in Bulk Earth ratios relative to calcium, and the
logarithmic abundance of calcium and helium [Ca/He] ranges from -7.0
to -12.0. This ratio is a parameter to be determined along with $T_\mathrm{eff}$
and $\log g$. The hydrogen abundance was fixed at [H/He] = -4.5.

Gaia EDR3 \citep{GaiaEDR3} reported parallaxes for 1473 of our objects, but only 1154 with parallax/error $\geq 1$. A further 181 have parallax in DR2 \citep{GaiaDR2}, but not EDR3.
We used the parallaxes and Gaia G magnitude to estimate the absolute magnitude in the G filter for these objects and used it to distinguish
between main sequence stars and subdwarf or white dwarf stars.
In our $\chi^2$ fitting procedure, we iterated on the combined observational
constraints of the SDSS spectrum, SDSS and Gaia photometry, and Gaia
parallaxes. The latter two provide a powerful constraint on the radius
and thus $\log g$, and also solve the common degeneracy between hot and
cool solutions from the Balmer line fitting. Because of this
constraint, we do not apply the 3D corrections of \citet{Tremblay13} on the result of the spectroscopic fit, since that would
imply a change in $\log~g$, thus violating
the parallax constraint.  Statistical errors were derived from the
$\chi^2$ fitting; the largest contribution for many objects comes from the
parallax error, since we used data with parallax errors up to 25\%,
corresponding to 0.37 dex in $\log g$.

\section{Results}

\begin{figure}
   \centering
   \includegraphics[width=\linewidth]{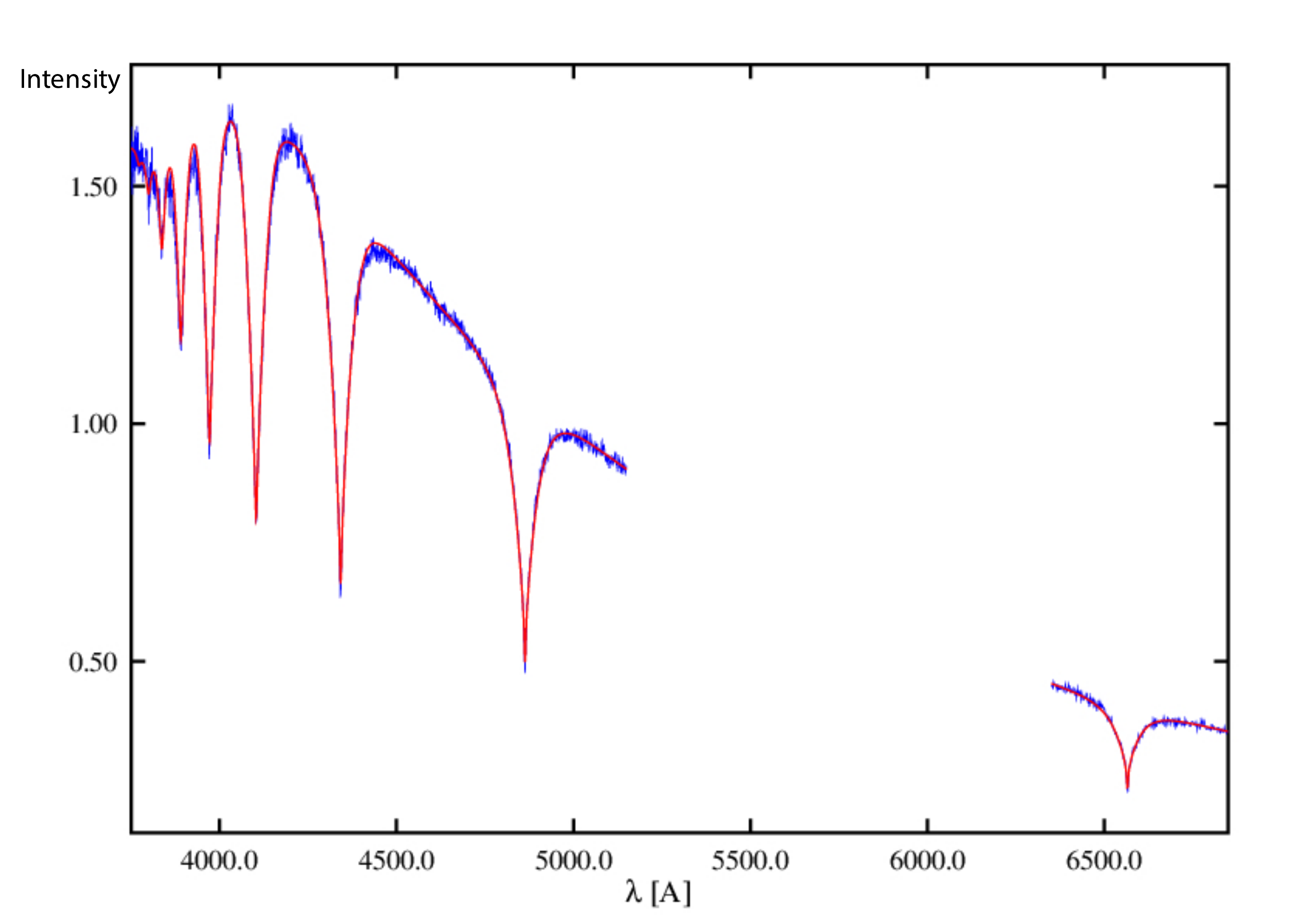} 
      \caption{DA SDSS~J075144.06+223004.80, g=16.639, P-M-F=11112-58428-0632 spectrum in blue, and in red the  best model fit at $T_\mathrm{eff}=20056 \pm 46$~K, $\log g=7.913\pm 0.025$, mass M=$0.579 \pm 0.017\,M_\odot$. Only the spectral regions used in our fitting routine are shown. A featureless region for DAs, between H$\beta$ and H$\alpha$, is not included in the fit.}
         \label{Fig:damodel}
   \end{figure}

\begin{figure}
   \centering
   \includegraphics[width=10cm]{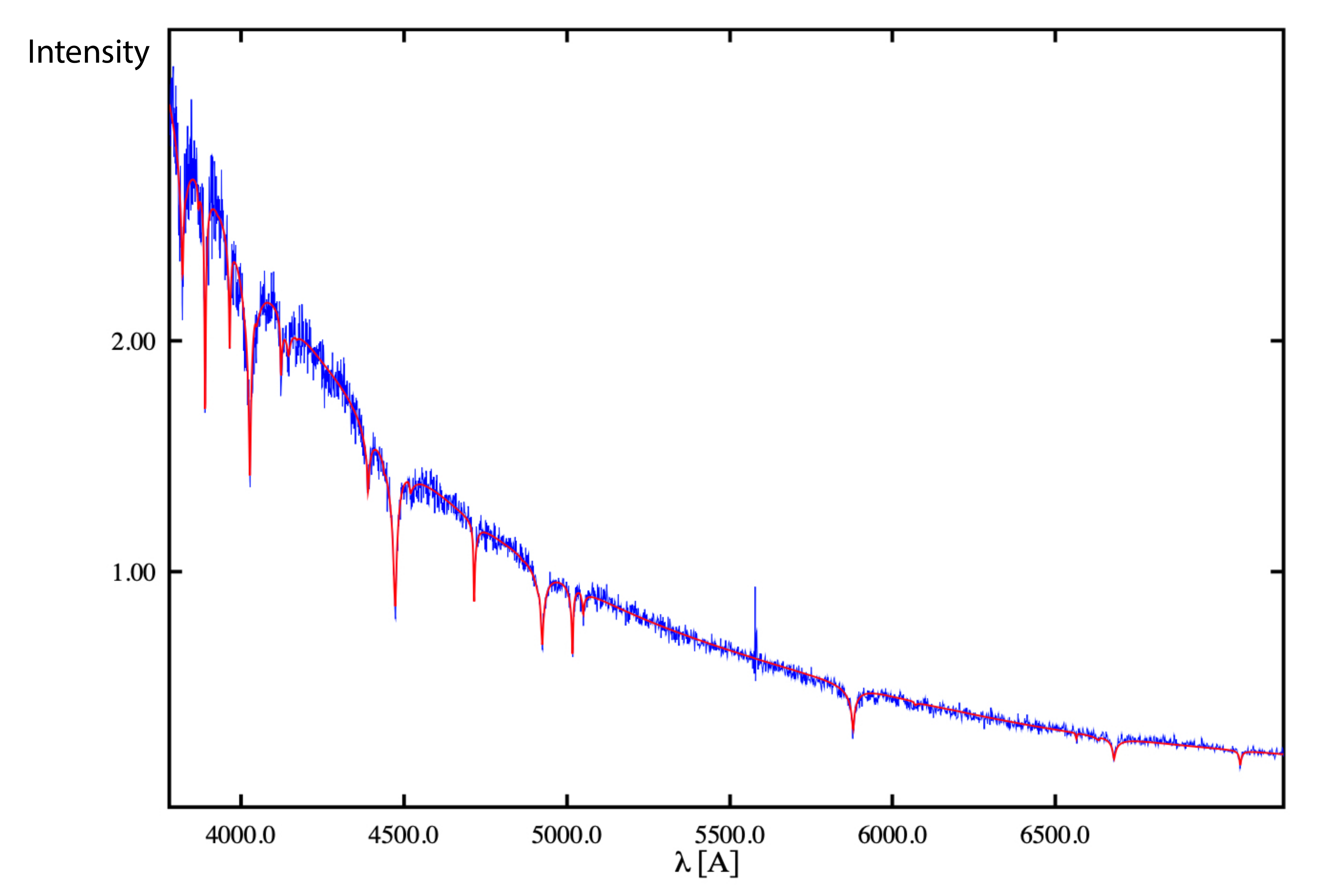}
    \caption{DB SDSS~J165222.17+322214.06, g=17.767, P-M-F=10909-58280-0663 spectrum in blue and best fitting DB model with $T_\mathrm{eff|}=16343 \pm 46$~K, $\log g=8.126 \pm 0.032$ in red.}
         \label{Fig:dbmodel}
   \end{figure}

The external uncertainties in our atmospheric parameters derived from spectral analysis are minimized by the use of only SDSS spectra, i.e., same telescope and only one spectrograph (BOSS), and fitting all the spectra with the same models and fitting technique.
Table~\ref{Table:da} shows the atmospheric parameters we obtained for DAs with (S/N)$_g\geq 10$ and parallax/uncertainty$\geq 4$. The complete table is available as supplementary material.
As an example of the fit, Fig~\ref{Fig:damodel}
shows the measured spectrum and the best fit model for one (S/N)$_g=70$, g=16.639 mag DA, while Fig.~\ref{Fig:dbmodel} shows the spectrum and a model for one selected DB. All the plots of the fits are made available online.

\begin{table}
    \centering
    \caption{Atmospheric parameters for DAs with (S/N)$_\geq 10$ and parallax/uncertainty$\geq 4$. The complete table is available in electronic form.\label{Table:da}}
    \tiny
    \begin{tabular}{lccccccccccccccc}
SDSS J             &P-M-F           &$T_\mathrm{eff}$&$\sigma_T$&$\log g$&$\sigma_{\log g}$&$V_r$(km/s)&$\sigma_V$&S/N&$\chi^2$&d(pc)&z(pc)&E(B-V)&$R_\lambda$&$M(M_\odot)$&$\sigma_M$\cr
000211.25+114243.06&11561-58485-0546&11832&116&8.125&0.178&+033&02&10.5&0.991&353&268&0.063&0.739&0.674&0.147\cr
000356.67+110603.64&11561-58485-0612&14065&261&8.916&0.105&+167&22&13.7&0.981&211&161&0.057&0.555&1.128&0.070\cr
002015.88+085724.05&11312-58433-0651&08412&061&8.152&0.225&+000&17&08.5&1.065&207&165&0.136&0.563&0.683&0.189\cr
002354.04+072540.33&11313-58426-0562&07940&198&8.463&0.231&+075&47&03.7&0.847&190&155&0.022&0.542&0.880&0.202\cr
004105.49+064152.32&11046-58398-0107&09415&030&8.021&0.063&+000&10&17.1&1.107&168&139&0.024&0.503&0.611&0.049\cr
011526.98+160445.56&11067-58507-0475&12855&205&7.950&0.032&+009&19&12.5&0.994&148&107&0.128&0.415&0.581&0.023\cr
012042.38+151322.53&11067-58507-0210&25278&301&8.145&0.122&+000&16&18.8&1.083&370&271&0.060&0.742&0.708&0.097\cr
012419.71+070726.42&11077-58433-0010&06545&120&7.682&0.374&+012&25&06.1&0.943&215&176&0.032&0.585&0.447&0.223\cr
012458.22+152152.86&11068-58488-0711&31173&273&8.535&0.186&+077&34&12.4&1.215&416&303&0.042&0.780&0.946&0.142\cr
012534.99+160603.50&11059-58515-0473&09755&075&8.644&0.169&+025&26&08.1&1.125&189&136&0.057&0.494&0.993&0.132\cr
012851.81+151311.02&11059-58515-0326&08528&122&7.691&0.029&-015&32&04.0&0.838&123&089&0.032&0.362&0.459&0.016\cr
013436.86+135226.76&11060-58523-0314&26720&322&8.038&0.132&+054&18&16.2&1.017&401&296&0.048&0.773&0.655&0.090\cr
013842.60+082449.66&11071-58429-0748&12267&093&8.085&0.169&+053&12&18.2&1.013&317&252&0.040&0.717&0.652&0.135\cr
014042.04+170048.17&11051-58510-0768&10671&075&8.340&0.195&+006&21&10.3&1.126&257&180&0.057&0.594&0.804&0.172\cr
014102.48+095850.16&11062-58509-0530&08971&077&8.056&0.311&+020&25&06.9&1.004&317&246&0.071&0.709&0.629&0.239\cr
014413.38+140719.39&11052-58438-0354&18275&121&8.003&0.119&+040&12&23.6&0.949&275&200&0.051&0.633&0.620&0.087\cr
014518.98+172343.60&11044-58508-0549&11982&124&8.291&0.168&+013&21&11.1&1.034&316&218&0.043&0.665&0.778&0.147\cr
015436.83+051656.71&11650-58508-0017&11061&455&8.516&0.229&-089&90&02.4&1.122&213&173&0.037&0.580&0.916&0.187\cr
015640.60+142813.13&11045-58485-0858&08708&050&8.122&0.213&+000&15&11.1&1.001&286&204&0.043&0.639&0.666&0.177\cr
072445.68+385044.86&09367-57758-0379&11059&048&7.931&0.097&+084&11&20.3&0.913&223&086&0.055&0.350&0.567&0.067\cr
072559.08+411245.75&10656-58163-0426&28411&143&8.255&0.138&+012&13&24.9&1.095&366&146&0.063&0.520&0.780&0.112\cr
072559.68+353836.26&09363-57742-0527&18596&107&7.669&0.074&+007&08&30.1&1.104&323&121&0.044&0.454&0.489&0.041\cr
072607.92+395112.05&09367-57758-0733&36766&256&7.998&0.135&-011&16&32.1&0.870&475&188&0.048&0.610&0.654&0.086\cr
072634.39+395440.04&09367-57758-0763&09731&029&8.418&0.062&+067&10&20.0&1.013&146&058&0.047&0.252&0.852&0.056\cr
072655.14+403027.89&09367-57758-0743&25416&235&7.854&0.225&+043&11&25.4&0.952&505&202&0.043&0.636&0.565&0.126\cr
072902.28+392442.17&09367-57758-0159&11586&081&7.544&0.120&+139&13&16.5&1.056&343&138&0.056&0.498&0.412&0.063\cr
073018.36+411320.42&10656-58163-0276&14194&094&7.792&0.020&+018&08&27.2&1.072&131&054&0.060&0.239&0.522&0.005\cr
073129.34+371444.87&09366-57746-0398&12898&084&8.598&0.034&+057&09&28.2&0.901&114&045&0.048&0.204&0.968&0.029\cr
073237.88+420454.33&10656-58163-0844&17608&044&8.469&0.015&+056&04&63.4&1.087&088&037&0.052&0.172&0.893&0.012\cr
073247.77+213346.81&11085-58462-0525&16899&092&8.117&0.078&-029&10&26.3&1.051&263&083&0.045&0.340&0.679&0.058\cr
073348.71+443733.50&10655-58172-0717&08032&019&7.962&0.044&+099&05&31.3&1.093&118&051&0.054&0.228&0.575&0.034\cr
073357.00+283123.83&10285-58083-0738&16633&186&8.357&0.101&+039&21&13.4&1.086&240&086&0.039&0.352&0.822&0.086\cr
073452.86+380301.56&09366-57746-0762&07592&033&8.085&0.054&+019&07&20.8&1.149&126&052&0.048&0.229&0.642&0.044\cr
073612.36+222937.20&11085-58462-0675&14785&125&8.463&0.027&+079&07&41.8&1.362&136&045&0.039&0.204&0.886&0.023\cr
073619.09+223338.89&11085-58462-0673&23016&131&8.008&0.067&+083&07&36.3&1.074&267&089&0.037&0.362&0.633&0.048\cr
073637.83+281324.25&10285-58083-0232&09859&120&7.717&0.104&+160&42&04.9&1.035&260&096&0.036&0.383&0.475&0.063\cr
073748.13+201835.57&11085-58462-0168&26510&232&8.277&0.141&+084&15&20.5&1.030&382&124&0.036&0.463&0.789&0.117\cr
073812.70+472941.35&10654-58429-0432&24044&110&8.118&0.035&+004&06&42.1&1.159&214&098&0.089&0.389&0.692&0.026\cr
    \end{tabular}
    \end{table}

For 68 DA+dM we fitted the blue part of the spectra to estimate the white dwarf properties, when ignoring the H$\alpha$ line was sufficient for a good fit.

\begin{figure}
   \centering
   \includegraphics[width=\linewidth]{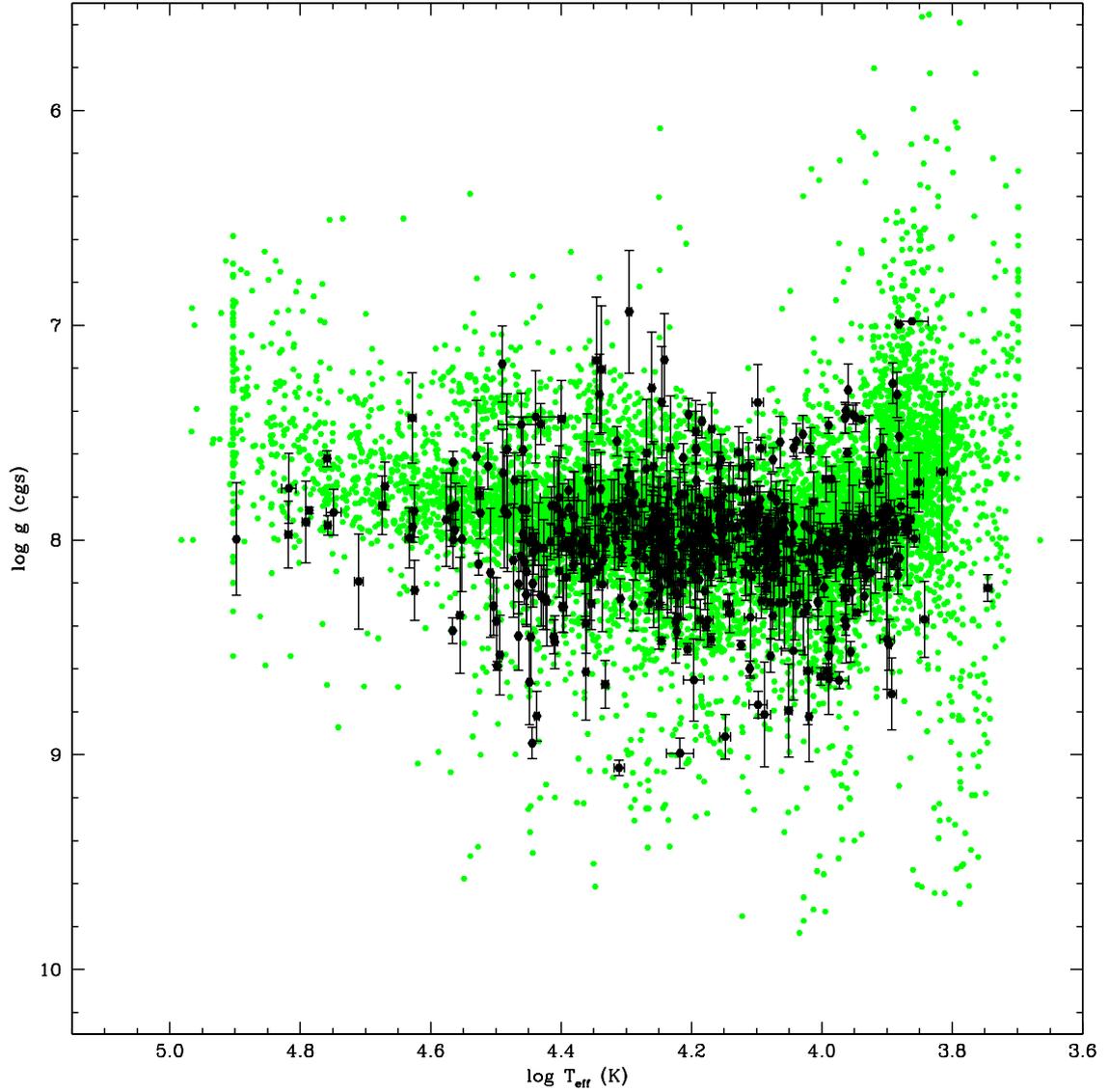}
      \caption{
      Surface gravity ($\log g$) and effective temperature ($T_{\rm eff}$) estimated for the 555 pure DA white dwarf stars for which the SDSS spectra has S/N$_g\geq 10$, and
Gaia EDR3 parallax/error$\geq 4$. In green, for comparison, the 11\,212 pure DAs in \citet{dr14} with S/N$_g\geq 10$.}
   \label{Fig:da}
\end{figure}

  \begin{figure}
   \centering
   \includegraphics[width=10cm]{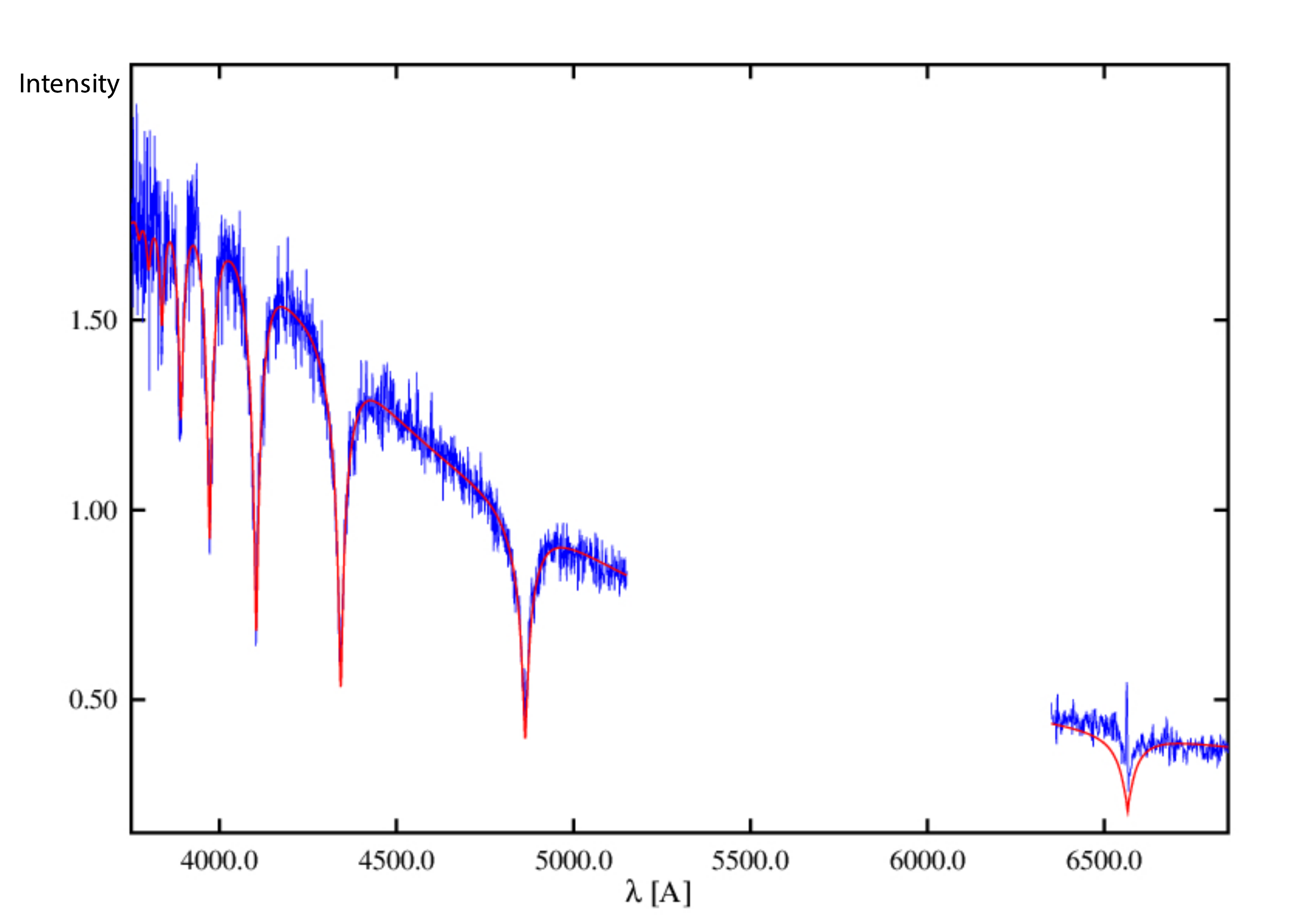}
    \includegraphics[width=10cm]{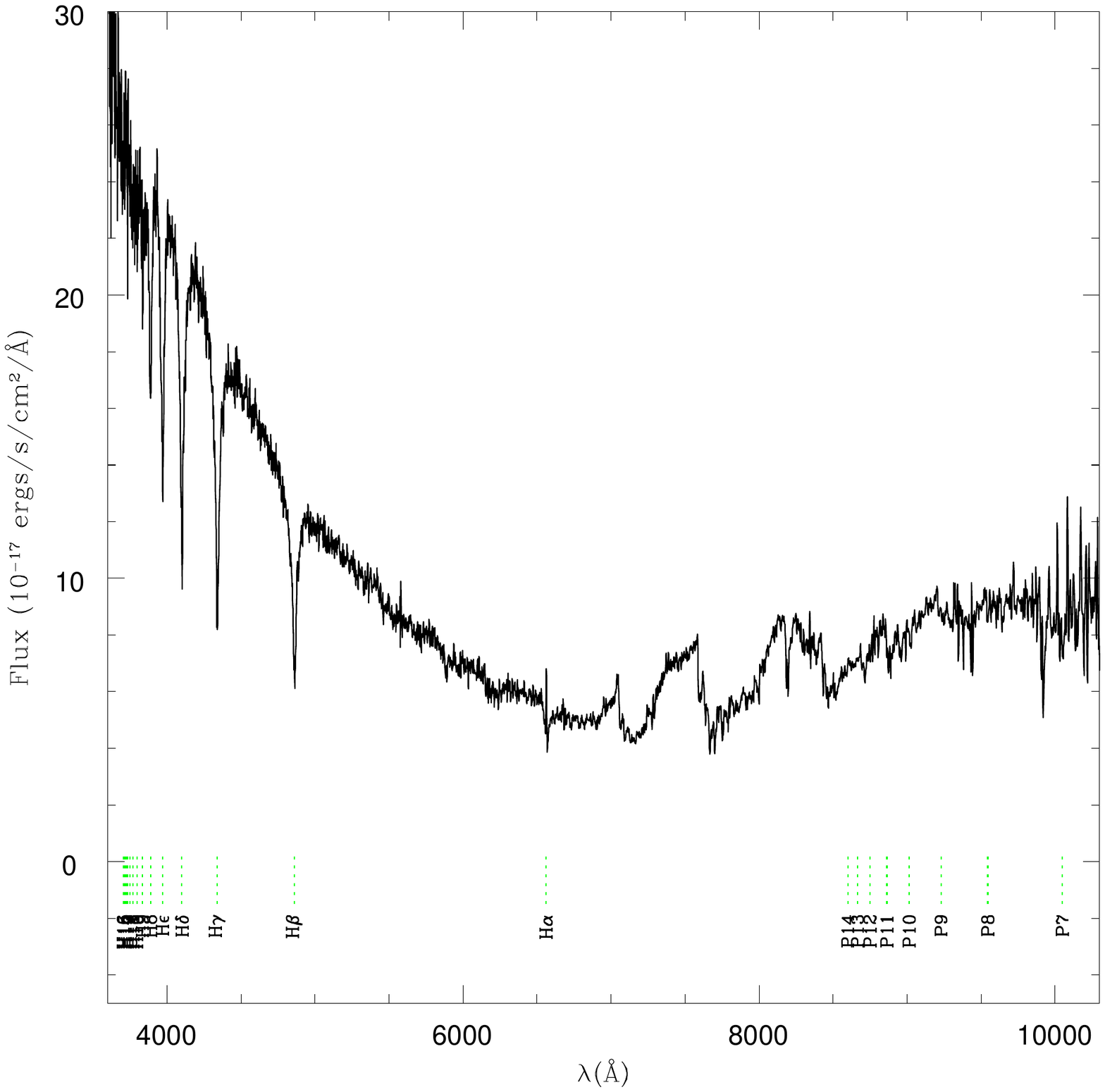}
    \caption{DA+M SDSSJ001626.39+104214.92, g=18.882, P-M-F=11565-58507-0320 spectrum in blue and best fitting DA model with $T_\mathrm{eff}=23586 \pm 272$~K, $\log g=7.561 \pm 0.037$ in red. H$\alpha$ 
      shows contamination from the M unresolved companion, and was not included in the fit.The bottom curve shows the whole SDSS spectrum.}
         \label{Fig:dammodel}
   \end{figure}

Table~\ref{tab:db} shows the parameters for the new 10 DBs with S/N$\geq 10$ following \citet{Koester15}.

\begin{table}
    \centering
     \caption{Atmospheric parameters for DBs}
    \label{tab:db}
    \begin{tabular}{l|c|c|r|c|c|r|r|r|c|r|}
Spectrum&Name&$T_\mathrm{eff}$&$\sigma_T$&$\log g$&$\sigma_{\log g}$&$V_r$(m/s)&$\sigma_V$&S/N&$M(M_\odot)$&$\sigma_M$\\ 
09176-58080-0181& SDSSJ2333+0051 & 19745 &  170 &  7.626 & 0.191  &    8 &  7 & 16.8&0.464&0.034\\
09355-57814-0682& SDSSJ0804+3513 & 11573 &   92 &  7.990 & 0.102  &   38 & 21 & 18.7&0.569&0.041\\
09369-58054-0103& SDSSJ0742+3906 & 14047 &   86 &  7.954 & 0.184  &   -8 & 10 & 14.9&0.554&0.065\\
10230-58224-0064& SDSSJ0957+3359 & 25659 &  476 &  8.045 & 0.075  &   31 &  9 & 18.7&0.629&0.029\\
10256-58193-0462& SDSSJ1337+3449 & 18714 &  118 &  8.063 & 0.088  &  -18 &  7 & 19.6&0.626&0.036\\
10909-58280-0663& SDSSJ1652+3222 & 16343 &   46 &  8.126 & 0.032  &    0 &  5 & 25.1&0.660&0.014\\
11123-58429-0222& SDSSJ0839+2553 & 12643 &   91 &  7.972 & 0.184  &  -26 & 12 & 16.5&0.561&0.067\\
11350-58455-0130& SDSSJ1010+2722 & 15258 &   69 &  7.630 & 0.297  &    0 &  8 & 13.4&0.457&0.063\\
11378-58437-0772& SDSSJ0944+3103 & 14293 &   85 &  7.981 & 0.239  &    0 & 10 & 13.6&0.570&0.085\\
11704-58514-0257& SDSSJ0938+2537 & 16436 &   60 &  8.040 & 0.179  &  -15 &  6 & 19.5&0.609&0.072\\
\end{tabular}
\end{table}

\subsection{Masses\label{section:masses}}

For white dwarfs, the main indicator of $\log~g$ is the width of the atmospheric absorption lines. However, for $T_\mathrm{eff} < 10\,000$~K, the width of the hydrogen lines becomes very weakly dependent on gravity. As a result, it is very difficult to distinguish low mass white dwarfs and metal-poor main sequence A/F stars in the $T_\mathrm{eff} < 10\,000$~K and $\log g < 6.5$ range solely with visual inspection, even though low metallicity main sequence stars have an upper limit to $\log g\lesssim 4.64$, for a turn-off mass of $\sim 0.85~M_{\odot}$. 
Following \citet{dr14}, the two steps we took to overcome this limitation were the extension of our pure H model grid to $\log g\geq 3.5$, fitting all the spectra
we visually classified as DAs and sdAs, using the result of the fit to separate $\log g \geq 6.5$ as white dwarfs, and finally, using the parallaxes from Gaia EDR3 to estimate the absolute magnitude
and use $M_G\geq 9.5$ as sdAs and white dwarfs, for those spectra showing only hydrogen lines.

At the cool end of our sample, $\log g=6.5$ corresponds to a mass around $0.2~M_\odot$, well below the single mass
evolution in the lifetime of the Universe --- but reachable via interacting binary evolution. The He-core white dwarf stars in the mass range $0.2-0.45~M_\odot$,
referred to as low-mass white dwarfs, are usually found in close binaries, often double degenerate systems \citep{Marsh95}, being most likely a product of interacting binary stars evolution.
More than 70\% of those studied by \citet{Brown11} with masses below $0.45~M_\odot$ and all but a few with masses below $0.3~M_\odot$ show radial velocity variations \citep{Brown13, Gianninas14,Brown17}.
\citet{Kilic2007} suggest that single low-mass white dwarfs result from the evolution of old metal-rich stars that truncate
evolution before the helium flash due to severe mass loss. They also conclude all white
dwarfs with masses below $\simeq 0.3~M_\odot$ must be a product of binary
star evolution involving interaction between the components.

\begin{figure}
   \centering
   \includegraphics[width=\linewidth]{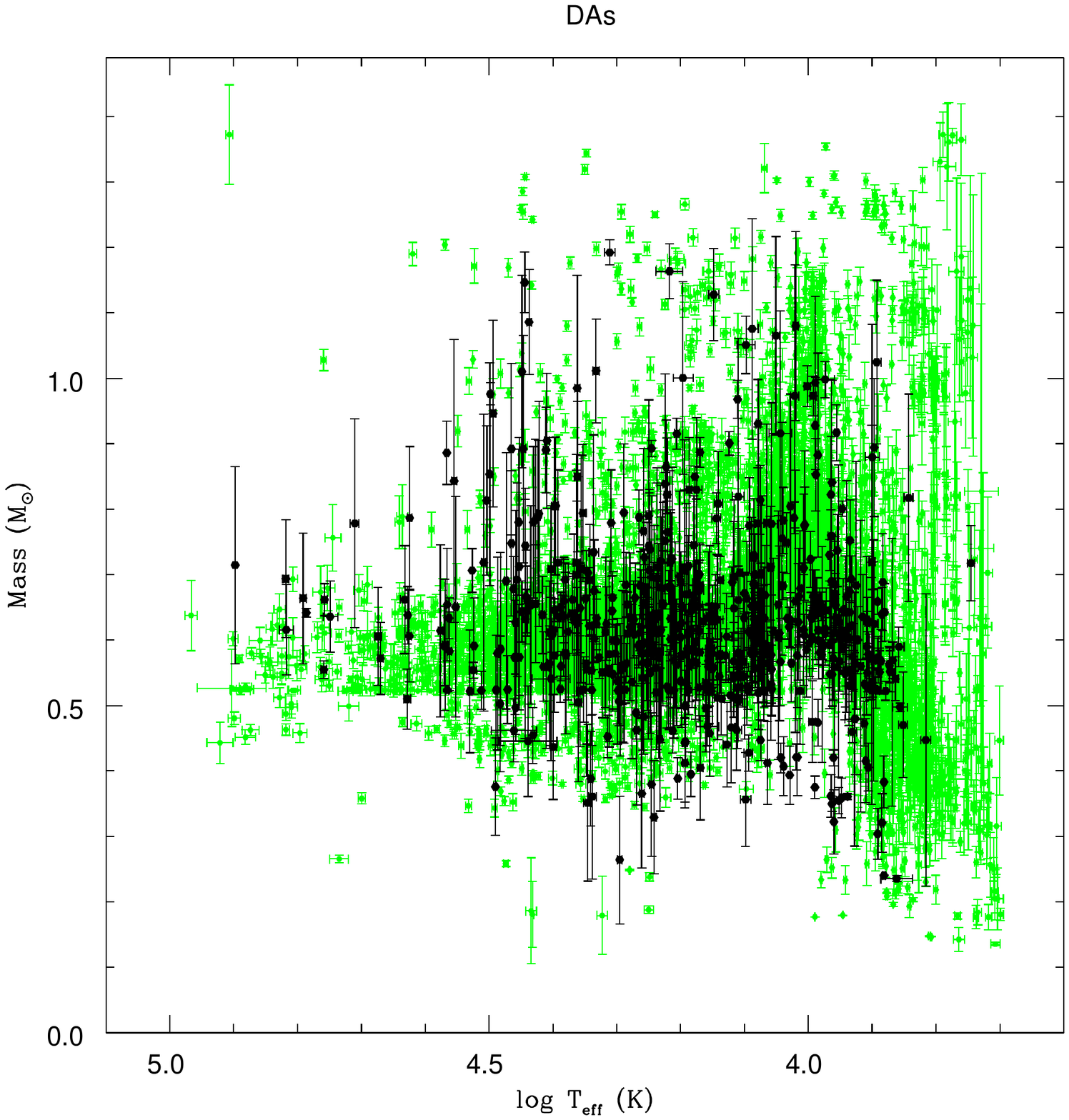}
      \caption{Masses for the 595 best DAs, estimated from the spectroscopic/distances parameters. In the background, we plot the 5951 pure DAs with S/N$_g\geq 15$ from \citet{dr14}, corrected to 3D.}
         \label{Fig:damass}
   \end{figure}

Table~\ref{Table:da} shows the atmospheric parameters obtained from the fitting of the spectra for DAs. The mean mass for these 595 DAs is $\langle M_\mathrm{DA} \rangle = 0.6283\pm 0.0056~M_\odot$, where the quoted uncertainty refers to the uncertainty in the mean value itself. The one sigma dispersion of the whole distribution is $0.260~M_\odot$. Fig.~\ref{Fig:damass} shows the masses obtained versus effective temperature. A comparison to the masses obtained in \citet{dr14} with the 3D corrections \citep{Tremblay13} is also shown in the same figure.

\citet{Tremblay19} fitted 3171 S/N$\geq 20$ SDSS DR14 spectra for DAs
white dwarfs selected by \citet{nicola18},
applying 3D corrections 
and compared to those they obtain from the {\it Gaia} photometry and parallax, concluding the agreement is good, for DAs, within 2\%. They obtained a mean mass $\langle M_\mathrm{DA}\rangle=0.586~M_\odot$, with a dispersion of $0.150~M_\odot$. For the 1145 DAs in \citet{Gianninas11} with good Gaia DR2 parallaxes they obtained a mean mass $\langle M_\mathrm{DA}\rangle=0.599~M_\odot$, with a dispersion of $0.165~M_\odot$. 
\citet{Genest19} concluded the photometric and spectroscopic analysis of 2236 DAs on their SDSS sample with Gaia DR2 parallaxes/error$\geq 10$ agreed within $1\sigma$ for 60.9\% of their sample and obtained
$\langle M_\mathrm{phot}^\mathrm{DA}\rangle = 0.617~M_\odot$, with a dispersion of $0.125~M_\odot$ and $\langle M_\mathrm{spec}^\mathrm{DA}\rangle = 0.615~M_\odot$, with a dispersion  of $0.147~M_\odot$.
\citet{dr14} analysed 11\,129 DAs up to DR14 with S/N$_g\geq$10, obtaining a mean mass $\langle M_\mathrm{DA} \rangle=0.5903\pm 0.0014~M_\odot$, and individual dispersion of $0.152~M_\odot$. For the 8171 DAs with $T_\mathrm{eff}\geq 10\,000$~K, the mean mass was $\langle M_\mathrm{DA} \rangle=0.6131\pm 0.0014~M_\odot$, with a dispersion $0.126~M_\odot$, while for those 2958 with $T_\mathrm{eff}<10\,000$~K, $\langle M_\mathrm{DA}\rangle=0.5276\pm 0.0035~M_\odot$ with a dispersion $0.174~M_\odot$. The DR16 sample is not large enough to allow a study of the mass distribution and  its dependency with temperature. A future work analysing the whole SDSS sample with a consistent method and parallax will be necessary.

\begin{figure}
   \centering
   \includegraphics[width=\linewidth]{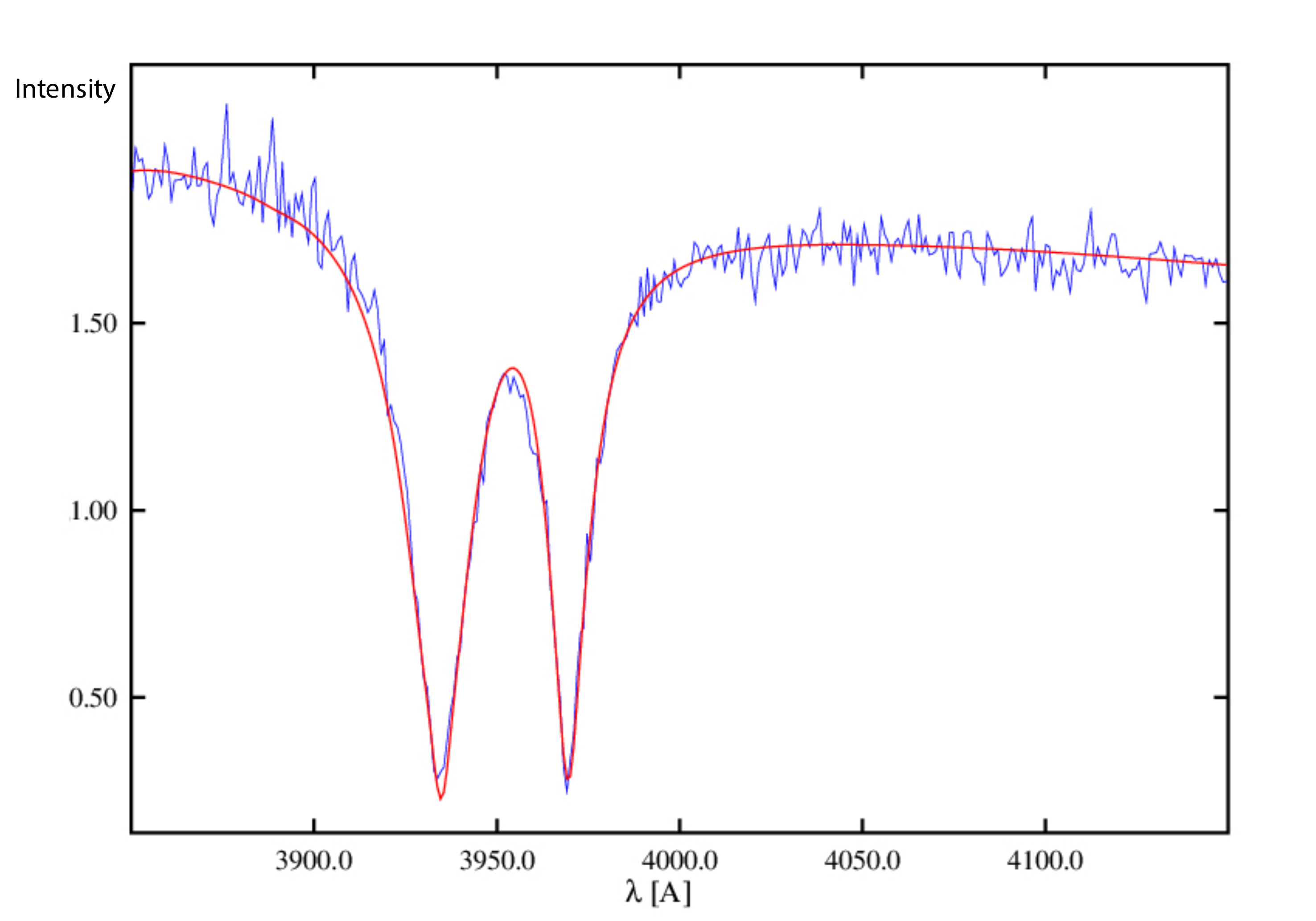}
      \caption{DZ SDSS~J131336.95+573800.5, g=16.895, P-M-F=1319-52791-0409 in blue, and in red the  best model fit at $T_\mathrm{eff}=8695 \pm 90$~K, $\log g=7.987\pm 0.022$, [Ca/He]=-9.194$\pm$0.010, mass=$0.559 \pm 0.009\,M_\odot$.}
         \label{Fig:dzmodel}
   \end{figure}

\begin{figure}
   \centering
   \includegraphics[width=\linewidth]{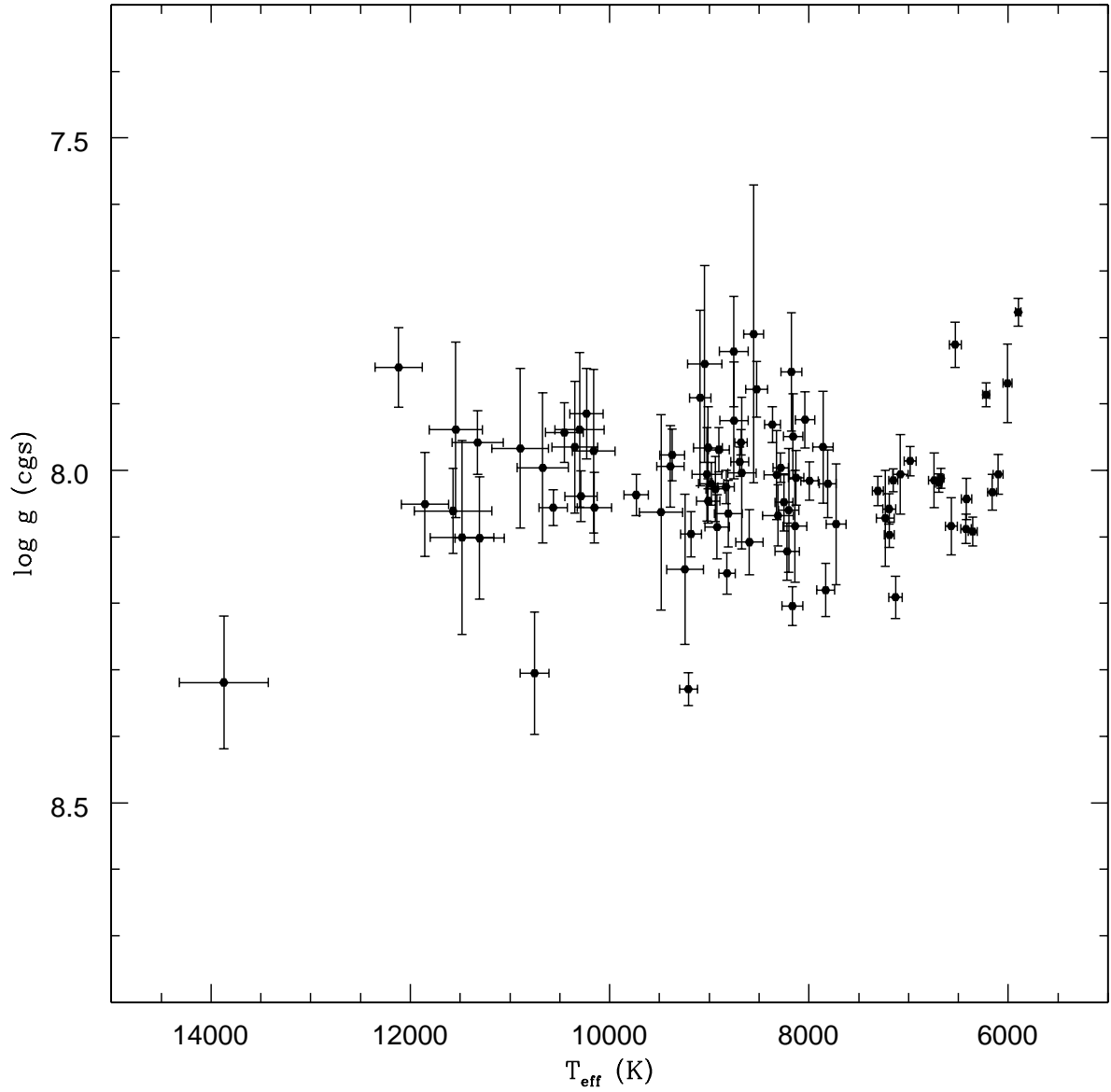}
      \caption{Atmospheric parameters for 85 DZs with S/N$_g\geq 20$ and Gaia parallaxes.}
         \label{Fig:dz}
   \end{figure}

With Gaia DR2 and EDR3 parallaxes, we were also able to fit simultaneously the photometry and spectra for 85 SDSS DZs with SN$_g\geq 20$ and parallax/error$\geq 4$, estimating their effective temperatures, surface gravities and [Ca/H] (see Fig.\ref{Fig:dz}. These objects were not restricted to DR16. In fact, only 3 DZs in DR16 have SN$_g\geq 20$, P-M-F=09161-57691-0879 SDSS~J212140.24+021737.32, 09174-58070-0242 SDSS~J231317.47+001201.16 and 10910-58254-0644 SDSS~J162625.86+351341.48, but none with parallax/error$\geq 4$. Fig.~\ref{Fig:dzmodel} shows one sample DZ spectra and our best fit model.

\begin{figure*}
   \centering
   \includegraphics[width=\linewidth]{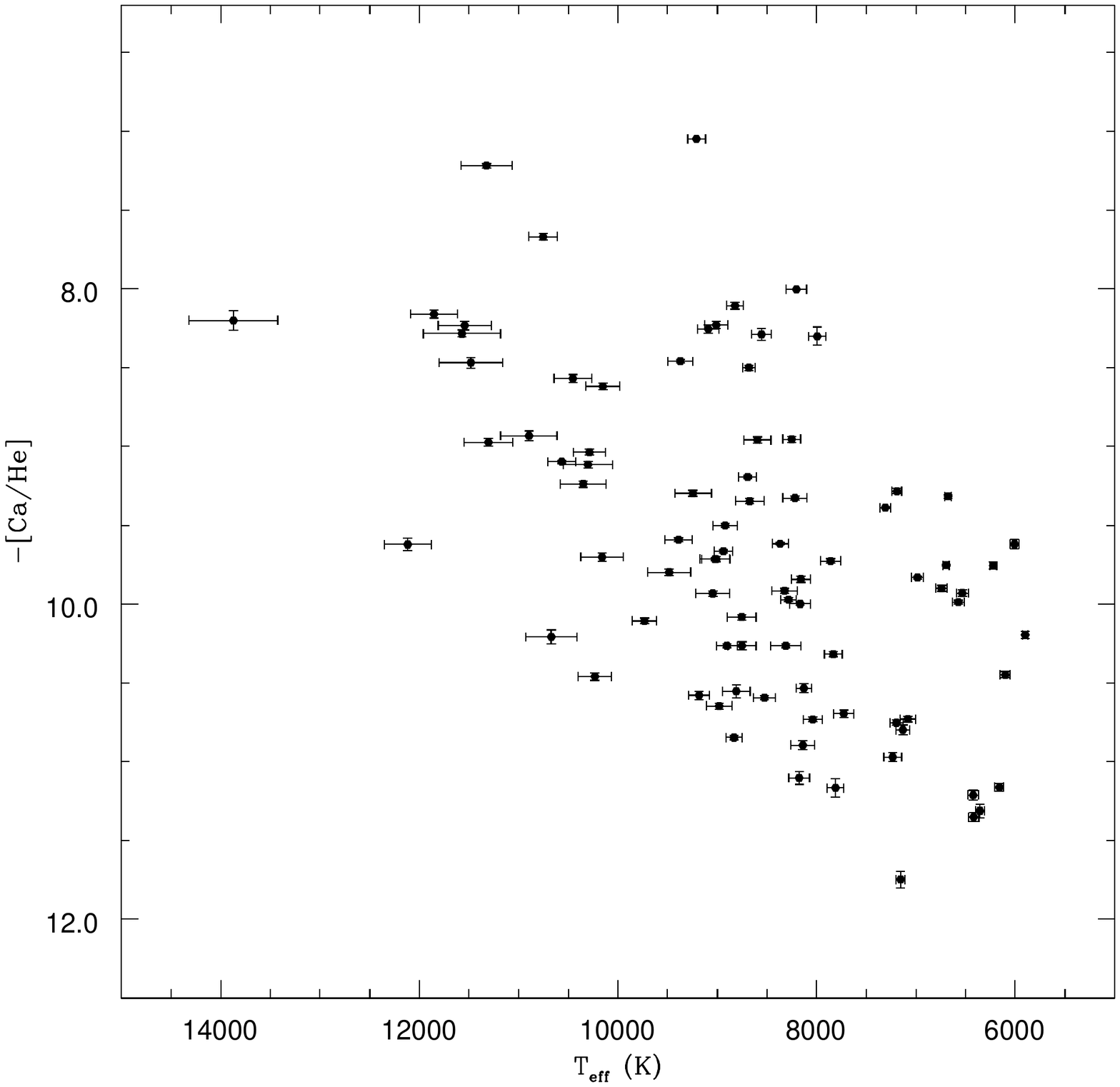}
      \caption{Ca determinations for the sample of S/N$_g\geq 20$ DZs.
      }
         \label{Fig:dzca}
   \end{figure*}

For DZs, we estimated their $T_\mathrm{eff}$ and $\log g$ from new atmospheric models.
The mean parameters for the 85 analysed DZs 
range from 13870~K to 5900~K and
7.762 to 8.329 dex are $\langle T_\mathrm{eff}^\mathrm{DZ} \rangle = 8685 \pm 1586$~K and $\langle \log g_\mathrm{DZ} \rangle = 8.017 \pm 0.011$~dex(cgs) (weighted mean 8.020~dex), and a mean mass of
$\langle M_\mathrm{DZ} \rangle = 0.640 \pm 0.013\,M_\odot$,
compared to the sample of 555 DAs, ranging from 79\,000~K$\leq T_\mathrm{eff}\leq$5\,600~K, 6.937$\leq \log g\leq$9.061~dex 
$\langle T_\mathrm{eff}^\mathrm{DA}\rangle =16803\pm 380$~K, $\langle \log g_\mathrm{DA} \rangle =7.998\pm 0.011$, and mean mass
$\langle M_\mathrm{DA} \rangle =0.618\pm 0.006\,M_\odot$, and a one sigma dispersion of $0.123~M_\odot$. Fig.~\ref{Fig:dzca} shows the calcium determinations for the sample. The sample reported is not large enough to allow a study of the mass distribution for a meaningful comparison with the DA or DB mean mass, and the dependency of the calcium abundances with temperature due to the deepening of the convection zone at lower temperatures.
\begin{table}
    \centering
    \caption{Parameters for the 85 DZs with (S/N)$_g \geq 20$ SDSS spectra and Gaia
    EDR3 parallaxes, for models with metals fixed at bulk Earth (B.E.) ratios to Ca, [Ca/He] from -7.00 to -15.00.}
    \label{tab:dz.fit}
     \tiny
    \begin{tabular}{lcccccccccccccccccc}
P-M-F&Name&$T_\mathrm{eff}$&$\sigma_T$&$\log g$&$\sigma_{\log g}$&m-M &$\sigma_{m-M}$&d(pc)&z(pc)&E(B-V)&$R_\lambda$&-[Ca/He]&$\sigma_{[Ca/He]}$&$V_r$&$\sigma_{V}$&S/N&M&$\sigma_M$\cr
0305-51613-0197&SDSSJ1425-0050&07131&068&8.191&0.032&4.249&0.031&0070&0057&0.042&0.249&10.799&0.031&012&33&14.1&0.887&0.111\cr
0335-52000-0247&SDSSJ1234-0330&08809&142&8.065&0.050&5.148&0.053&0107&0091&0.031&0.368&10.554&0.040&026&17&16.0&0.608&0.021\cr
0447-51877-0499&SDSSJ0848+5214&08130&077&8.011&0.041&5.128&0.050&0106&0066&0.021&0.283&10.535&0.029&094&15&15.1&0.573&0.018\cr
0655-52162-0300&SDSSJ0036-1112&07192&050&8.097&0.019&3.594&0.012&0052&0050&0.026&0.222&09.284&0.017&159&13&16.1&0.625&0.129\cr
0660-52177-0277&SDSSJ0113-0959&10455&189&7.943&0.045&5.568&0.038&0129&0123&0.031&0.461&08.567&0.025&044&06&16.5&0.539&0.017\cr
0737-52518-0009&SDSSJ2228+1207&06696&032&8.019&0.014&2.653&0.006&0033&0020&0.074&0.098&09.753&0.007&116&06&23.7&0.879&0.001\cr
0934-52672-0553&SDSSJ0848+3548&07832&089&8.180&0.040&4.753&0.045&0089&0055&0.027&0.242&10.318&0.017&041&14&16.1&0.678&0.018\cr
1010-52649-0084&SDSSJ1049+5154&06532&060&7.811&0.034&4.504&0.028&0079&0066&0.008&0.282&09.930&0.021&130&19&10.1&0.872&0.001\cr
1281-52753-0439&SDSSJ1309+4913&08284&078&7.996&0.022&4.484&0.015&0078&0072&0.006&0.305&09.972&0.013&048&07&19.3&0.564&0.009\cr
1319-52791-0409&SDSSJ1313+5738&08694&090&7.987&0.022&4.162&0.008&0067&0058&0.010&0.253&09.194&0.010&000&04&26.2&0.559&0.009\cr
1345-52814-0032&SDSSJ1351+4253&06424&051&8.088&0.022&2.977&0.006&0039&0037&0.008&0.169&11.212&0.031&089&44&18.6&0.882&0.001\cr
1655-53523-0127&SDSSJ1558+2512&07196&065&8.058&0.023&3.663&0.010&0054&0040&0.056&0.183&10.753&0.018&129&17&20.4&0.601&0.140\cr
1691-53260-0265&SDSSJ1647+2636&06984&060&7.986&0.022&3.518&0.007&0050&0030&0.074&0.143&09.830&0.010&090&09&18.4&0.878&0.001\cr
1694-53472-0154&SDSSJ1246+1155&07995&088&8.016&0.029&4.513&0.023&0079&0077&0.023&0.320&08.301&0.058&090&16&13.6&0.576&0.012\cr
2037-53446-0579&SDSSJ1125+3823&10154&171&8.056&0.053&5.590&0.063&0131&0122&0.016&0.458&08.618&0.021&094&07&17.0&0.605&0.022\cr
2170-53875-0154&SDSSJ1554+1735&05900&029&7.762&0.021&3.753&0.016&0056&0041&0.030&0.186&10.196&0.022&166&27&09.3&0.871&0.001\cr
2301-53712-0599&SDSSJ0618+6413&08753&142&7.925&0.088&6.353&0.106&0186&0066&0.090&0.282&10.264&0.024&025&11&15.4&0.524&0.030\cr
2367-53763-0546&SDSSJ1034+2245&06676&032&8.012&0.015&3.053&0.009&0040&0034&0.017&0.160&09.317&0.012&130&08&22.9&0.879&0.001\cr
2406-54084-0328&SDSSJ1005+2655&07859&104&7.965&0.084&5.756&0.105&0141&0113&0.025&0.432&09.726&0.015&088&10&16.2&0.545&0.032\cr
2421-54153-0336&SDSSJ0809+1112&06356&046&8.092&0.022&3.602&0.016&0052&0020&0.029&0.095&11.312&0.044&048&77&12.5&0.882&0.001\cr
2512-53877-0570&SDSSJ1156+1822&07154&044&8.015&0.017&2.133&0.004&0026&0025&0.027&0.121&11.748&0.052&092&46&42.3&0.580&0.149\cr
2629-54087-0330&SDSSJ2305+2307&11853&239&8.051&0.078&6.477&0.104&0197&0109&0.171&0.421&08.160&0.024&-06&04&25.9&0.606&0.033\cr
2907-54580-0126&SDSSJ1428+4403&06221&033&7.886&0.018&3.261&0.006&0044&0040&0.011&0.183&09.756&0.022&095&22&10.8&0.874&0.001\cr
2922-54612-0509&SDSSJ1244-0118&08203&104&8.060&0.093&6.126&0.124&0167&0147&0.029&0.522&08.002&0.001&-46&01&09.6&0.603&0.041\cr
3236-54892-0599&SDSSJ1258+3047&07233&088&8.072&0.072&5.343&0.094&0117&0116&0.010&0.442&10.971&0.027&095&36&14.8&0.610&0.139\cr
3251-54882-0406&SDSSJ1046+2424&06160&045&8.033&0.027&3.858&0.024&0059&0052&0.030&0.229&11.162&0.025&075&38&18.5&0.879&0.001\cr
3428-54979-0618&SDSSJ1616+4600&06100&050&8.006&0.030&4.613&0.026&0083&0059&0.010&0.258&10.449&0.020&083&24&11.3&0.878&0.001\cr
3480-54999-0359&SDSSJ1640+3154&06743&057&8.015&0.041&5.239&0.050&0111&0072&0.021&0.304&09.899&0.019&117&19&10.4&0.879&0.001\cr
3609-55201-0385&SDSSJ0201-0039&09373&123&7.977&0.039&5.229&0.039&0111&0094&0.024&0.378&08.459&0.013&053&04&25.9&0.555&0.016\cr
3752-55236-0550&SDSSJ0749+3124&06572&059&8.084&0.043&4.806&0.051&0091&0039&0.050&0.178&09.988&0.012&092&10&15.3&0.882&0.001\cr
3775-55207-0079&SDSSJ1143-0145&09046&172&7.840&0.148&6.581&0.189&0207&0173&0.014&0.580&09.933&0.017&097&06&19.1&0.494&0.038\cr
3814-55535-0008&SDSSJ0900+0331&07810&085&8.020&0.051&5.326&0.064&0116&0058&0.052&0.253&11.167&0.058&104&38&16.7&0.578&0.022\cr
3818-55532-0016&SDSSJ0909-0045&08674&142&8.004&0.114&6.475&0.158&0197&0098&0.027&0.388&09.346&0.016&115&08&16.6&0.570&0.046\cr
3835-55570-0386&SDSSJ1049-0007&09092&107&7.891&0.132&6.512&0.165&0200&0153&0.036&0.536&08.255&0.027&059&11&14.3&0.510&0.040\cr
3840-55574-0125&SDSSJ1137-0023&09484&215&8.063&0.147&6.650&0.204&0213&0179&0.017&0.593&09.798&0.021&000&07&17.1&0.608&0.062\cr
3861-55274-0690&SDSSJ1353+3239&08039&098&7.924&0.042&5.263&0.037&0112&0109&0.011&0.421&10.732&0.014&034&09&28.8&0.522&0.014\cr
3955-55678-0456&SDSSJ1516+2118&09183&103&8.096&0.034&5.158&0.036&0107&0089&0.045&0.362&10.579&0.027&026&10&27.5&0.627&0.015\cr
4007-55327-0548&SDSSJ1307+0307&08159&098&7.949&0.064&5.646&0.075&0134&0122&0.023&0.459&09.842&0.019&051&11&12.7&0.536&0.024\cr
4035-55383-0874&SDSSJ1414-0113&08525&111&7.878&0.042&5.175&0.038&0108&0089&0.046&0.359&10.594&0.013&025&05&35.9&0.505&0.010\cr
4065-55368-0194&SDSSJ1643+1422&06420&053&8.043&0.031&4.395&0.031&0075&0043&0.060&0.194&11.352&0.028&262&48&20.1&0.880&0.001\cr
4180-55679-0030&SDSSJ1703+2541&08904&105&7.969&0.033&5.199&0.028&0109&0061&0.031&0.265&10.265&0.013&029&06&26.7&0.549&0.013\cr
4218-55479-0010&SDSSJ0018-0012&09390&138&7.994&0.061&5.900&0.076&0151&0133&0.023&0.487&09.592&0.013&041&04&25.9&0.565&0.025\cr
4226-55475-0310&SDSSJ0106-0103&11545&266&7.939&0.132&6.778&0.171&0226&0203&0.054&0.638&08.233&0.026&-01&04&20.3&0.539&0.046\cr
4238-55455-0226&SDSSJ0229-0041&08140&119&8.084&0.084&5.954&0.113&0155&0126&0.029&0.469&10.895&0.029&023&26&17.0&0.618&0.037\cr
4395-55828-0220&SDSSJ0218-0919&09733&123&8.037&0.031&4.915&0.025&0096&0085&0.023&0.348&10.106&0.016&085&05&30.1&0.592&0.013\cr
4483-55587-0348&SDSSJ0823+2015&10302&247&7.939&0.116&6.716&0.142&0220&0106&0.034&0.414&09.116&0.020&049&04&20.8&0.536&0.040\cr
4489-55545-0877&SDSSJ0834+1846&08175&105&7.852&0.089&5.828&0.108&0146&0074&0.027&0.313&11.103&0.042&066&20&20.3&0.496&0.021\cr
4543-55888-0238&SDSSJ0046+0704&09024&151&8.006&0.070&5.762&0.088&0142&0117&0.038&0.444&09.714&0.010&036&04&29.3&0.572&0.029\cr
4567-55589-0822&SDSSJ1010+3948&08250&088&8.048&0.043&5.447&0.052&0122&0099&0.010&0.393&08.954&0.018&079&08&19.4&0.596&0.018\cr
4631-55617-0698&SDSSJ1055+3509&10349&231&7.965&0.099&6.478&0.123&0197&0177&0.016&0.589&09.238&0.020&010&04&22.4&0.551&0.037\cr
4636-55945-0276&SDSSJ0951+4033&08166&105&8.204&0.029&4.297&0.019&0072&0055&0.009&0.244&09.999&0.009&058&06&24.9&0.693&0.013\cr
4691-55651-0006&SDSSJ1037+4341&10897&285&7.967&0.120&6.930&0.153&0243&0207&0.012&0.645&08.932&0.032&053&06&15.8&0.554&0.044\cr
4732-55648-0248&SDSSJ1056+0128&10288&163&8.039&0.038&5.373&0.036&0118&0093&0.028&0.375&09.036&0.018&034&04&22.7&0.595&0.016\cr
4767-55946-0567&SDSSJ1133+0610&10232&168&7.915&0.068&6.002&0.078&0158&0140&0.027&0.504&10.460&0.023&044&09&28.1&0.523&0.022\cr
4780-55682-0354&SDSSJ1432+0354&11482&319&8.101&0.146&6.886&0.205&0238&0198&0.029&0.630&08.468&0.033&-04&06&15.8&0.635&0.063\cr
5112-55895-0838&SDSSJ0152+2418&07308&056&8.031&0.022&3.616&0.014&0052&0031&0.112&0.146&09.389&0.007&161&04&33.5&0.584&0.010\cr
5210-56003-0448&SDSSJ1547+0659&08367&079&7.931&0.027&4.716&0.022&0087&0060&0.029&0.261&09.616&0.008&083&04&27.5&0.526&0.010\cr
5314-55952-0385&SDSSJ0934+0822&08939&096&8.028&0.049&5.219&0.063&0110&0070&0.034&0.299&09.665&0.011&096&04&25.5&0.585&0.021\cr
5344-55924-0684&SDSSJ1031+0936&10672&258&7.996&0.113&6.516&0.151&0200&0159&0.030&0.550&10.208&0.043&040&13&17.5&0.570&0.045\cr
5472-55976-0418&SDSSJ1441+0831&07083&077&8.006&0.060&5.019&0.078&0100&0085&0.025&0.348&10.730&0.017&104&16&19.3&0.878&0.156\cr
5710-56658-0804&SDSSJ0939+5550&08681&061&7.958&0.019&4.176&0.012&0068&0048&0.018&0.216&08.500&0.017&057&05&26.7&0.542&0.008\cr
5720-56602-0356&SDSSJ0940+6136&08823&082&8.155&0.031&5.281&0.033&0113&0078&0.031&0.323&08.106&0.022&059&09&19.2&0.664&0.014\cr
5777-56280-0748&SDSSJ0917+2630&11305&246&8.102&0.092&6.329&0.125&0184&0124&0.029&0.464&08.974&0.024&048&04&22.9&0.635&0.039\cr
5875-56038-0566&SDSSJ1044+2023&08311&156&8.068&0.046&4.751&0.039&0089&0077&0.024&0.321&10.264&0.012&094&07&23.1&0.609&0.020\cr
5880-56042-0834&SDSSJ1143+1928&07728&100&8.081&0.091&5.321&0.125&0115&0110&0.018&0.425&10.696&0.024&084&18&17.0&0.616&0.040\cr
5945-56213-0334&SDSSJ0801+5329&09209&090&8.329&0.025&4.579&0.024&0082&0043&0.032&0.195&07.049&0.000&000&00&30.1&0.773&0.011\cr
6002-56104-0590&SDSSJ1339+2643&06008&046&7.869&0.059&5.067&0.069&0103&0101&0.007&0.398&09.619&0.029&168&33&07.8&0.874&0.001\cr
6008-56093-0381&SDSSJ1402+2506&08598&136&8.108&0.049&5.601&0.054&0131&0126&0.013&0.469&08.959&0.020&-19&09&16.1&0.633&0.022\cr
6009-56313-0790&SDSSJ1401+2840&08754&144&7.821&0.083&6.275&0.099&0179&0173&0.014&0.580&10.083&0.018&037&07&17.4&0.488&0.017\cr
6054-56089-0814&SDSSJ1506+4152&08981&129&8.020&0.032&4.904&0.019&0095&0081&0.015&0.336&10.647&0.019&016&08&31.8&0.580&0.013\cr
6153-56164-0216&SDSSJ2351+0633&09244&184&8.149&0.113&6.209&0.154&0174&0139&0.074&0.503&09.297&0.019&029&08&15.5&0.661&0.051\cr
6418-56354-0985&SDSSJ1104+2439&10753&144&8.305&0.092&6.088&0.146&0165&0150&0.011&0.529&07.669&0.021&028&08&16.0&0.761&0.041\cr
6467-56270-0724&SDSSJ0959+2556&08831&082&8.025&0.025&3.784&0.020&0057&0044&0.024&0.201&10.847&0.013&057&08&45.4&0.583&0.011\cr
6489-56329-0506&SDSSJ1319+3025&09012&145&7.966&0.062&6.027&0.071&0160&0159&0.010&0.549&09.715&0.015&093&06&20.1&0.548&0.025\cr
6639-56385-0860&SDSSJ1158+4252&12118&236&7.845&0.060&6.446&0.064&0194&0183&0.017&0.601&09.622&0.038&035&08&23.4&0.502&0.013\cr
6659-56607-0254&SDSSJ1027+4532&08555&099&7.795&0.224&6.379&0.283&0188&0156&0.015&0.543&08.288&0.037&049&12&13.5&0.481&0.061\cr
6674-56416-0868&SDSSJ1242+5226&11325&257&7.958&0.048&6.028&0.036&0160&0145&0.015&0.516&07.218&0.014&032&04&23.4&0.550&0.019\cr
6688-56412-0156&SDSSJ1212+5409&08219&123&8.122&0.043&5.462&0.044&0123&0109&0.014&0.421&09.327&0.015&092&08&16.1&0.641&0.019\cr
6698-56637-0260&SDSSJ1126+5241&08922&121&8.085&0.048&5.732&0.058&0140&0121&0.010&0.454&09.502&0.012&008&05&20.3&0.620&0.021\cr
6698-56637-0433&SDSSJ1120+5257&10566&141&8.056&0.027&4.529&0.015&0080&0069&0.016&0.292&09.097&0.010&017&02&39.0&0.606&0.011\cr
6795-56425-0480&SDSSJ1501+5609&09011&118&8.046&0.033&5.103&0.025&0104&0083&0.012&0.341&08.228&0.022&044&09&18.5&0.596&0.014\cr
6832-56426-0620&SDSSJ1234+5606&11572&390&8.061&0.064&6.202&0.045&0173&0151&0.010&0.532&08.284&0.022&028&04&22.8&0.612&0.027\cr
6974-56442-0808&SDSSJ1217+6419&10161&213&7.971&0.123&6.947&0.161&0245&0194&0.015&0.621&09.702&0.026&008&06&17.0&0.554&0.046\cr
7053-56564-0410&SDSSJ0241-0533&08323&129&8.007&0.067&5.781&0.085&0143&0118&0.025&0.448&09.916&0.017&069&08&14.8&0.571&0.029\cr
7253-56598-0339&SDSSJ0356-0631&13871&445&8.319&0.100&6.816&0.156&0230&0153&0.066&0.535&08.200&0.063&024&09&14.7&0.773&0.045\cr
    \end{tabular}
  
\end{table}

\section{Conclusions}
We extended our search for new spectroscopically confirmed white dwarf and subdwarf stars to SDSS DR16. 
The SDSS flux calibration is based on hundreds of comparison stars and in general more accurate than those derived from single night observations.
We fit the spectra of the highest signal-to-noise for each star, taking into account that SDSS re-observes fields and improves the quality of the spectra.

The total number of unique stars in \citet{dr7,dr10,dr12,dr14} and this paper is 30\,086 DAs, 2390 DCs, 2160 DBs, 1316 DZs, 572 DQs, 137 DOs, 4 DS, 396 sdB, 410 sdOs, and 363 CVs, i.e., DAs correspond to 82\% of the white dwarfs with SDSS spectra, excluding subdwarfs and CVs.

After the conclusion of our work, \citet{nicola21} submitted a visual classification of 998 DR16 spectra, 777 in common with our sample, and our classifications agree.
Their remaining 221 spectra are either re-observations of already classified white dwarfs or not white dwarfs.

\section{Data Availability}
The data underlying this article are available in the article and in its online supplementary material. 
The spectra are available on the Sloan Digital Sky Survey databases. 

\section*{Acknowledgements}
This study was financed in part by the Coordena\c{c}\~ao de Aperfei\c{c}oamento de Pessoal de N\'{\i}vel Superior - Brasil (CAPES) - Finance Code 001, Conselho Nacional de Desenvolvimento Cient\'{\i}fico e Tecnol\'ogico - Brasil (CNPq), and Funda\c{c}\~ao de Amparo \`a Pesquisa do Rio Grande do Sul (FAPERGS) - Brasil. IP acknowledges support from the UK's Science and Technology Facilities Council (STFC), grant ST/T000406/1. Funding for the Sloan Digital Sky Survey IV has been provided by the Alfred P. Sloan Foundation, the U.S. Department of Energy Office of Science, and the Participating Institutions. SDSS-IV acknowledges
support and resources from the Center for High-Performance Computing at
the University of Utah. The SDSS website is www.sdss.org.
SDSS-IV is managed by the Astrophysical Research Consortium for the 
Participating Institutions of the SDSS Collaboration including the 
Brazilian Participation Group, the Carnegie Institution for Science, 
Carnegie Mellon University, the Chilean Participation Group, the French Participation Group, Harvard-Smithsonian Center for Astrophysics, 
Instituto de Astrof\'{\i}sica de Canarias, The Johns Hopkins University, 
Kavli Institute for the Physics and Mathematics of the Universe (IPMU) / 
University of Tokyo, Lawrence Berkeley National Laboratory, 
Leibniz Institut f\"ur Astrophysik Potsdam (AIP),  
Max-Planck-Institut f\"ur Astronomie (MPIA Heidelberg), 
Max-Planck-Institut f\"ur Astrophysik (MPA Garching), 
Max-Planck-Institut f\"ur Extraterrestrische Physik (MPE), 
National Astronomical Observatories of China, New Mexico State University, 
New York University, University of Notre Dame, 
Observat\'ario Nacional / MCTI, The Ohio State University, 
Pennsylvania State University, Shanghai Astronomical Observatory, 
United Kingdom Participation Group,
Universidad Nacional Aut\'onoma de M\'exico, University of Arizona, 
University of Colorado Boulder, University of Oxford, University of Portsmouth, 
University of Utah, University of Virginia, University of Washington, University of Wisconsin, 
Vanderbilt University, and Yale University.

This research has made use of NASA's Astrophysics Data System Bibliographic Services,
SIMBAD database, operated at CDS, Strasbourg, France, and 
IRAF, distributed by the National Optical Astronomy Observatory, which is operated by the Association of Universities for Research in Astronomy (AURA) under a cooperative agreement with the National Science Foundation.
This work presents results from the European Space Agency (ESA) space mission Gaia. Gaia data are being processed by the Gaia Data Processing and Analysis Consortium (DPAC). Funding for the DPAC is provided by national institutions, in particular the institutions participating in the Gaia MultiLateral Agreement (MLA). The Gaia mission website is https://www.cosmos.esa.int/gaia. The Gaia archive website is https://archives.esac.esa.int/gaia.

The Gaia mission and data processing have financially been supported by, in alphabetical order by country:
the Algerian Centre de Recherche en Astronomie, Astrophysique et G\'eophysique of Bouzareah Observatory;
the Austrian Fonds zur F\"orderung der wissenschaftlichen Forschung (FWF) Hertha Firnberg Programme through grants T359, P20046, and P23737;
the BELgian federal Science Policy Office (BELSPO) through various PROgramme de D'eveloppement d'Exp\'eriences scientifiques (PRODEX) grants and the Polish Academy of Sciences - Fonds Wetenschappelijk Onderzoek through grant VS.091.16N;
the Brazil-France exchange programmes Funda\c{c}\~ao de Amparo \`a Pesquisa do Estado de S~ao Paulo (FAPESP) and Coordena\c{c}\~ao de Aperfeicoamento de Pessoal de N\'{\i}vel Superior (CAPES) - Comit\'e Fran\c{c}ais d'Evaluation de la Coop\'eration Universitaire et Scientifique avec le Br\'esil (COFECUB);
the Chilean Direcci\'on de Gesti\'on de la Investigaci\'on (DGI) at the University of Antofagasta and the Comit\'e Mixto ESO-Chile;
the National Science Foundation of China (NSFC) through grants 11573054 and 11703065;
the Czech-Republic Ministry of Education, Youth, and Sports through grant LG 15010, the Czech Space Office through ESA PECS contract 98058, and Charles University Prague through grant PRIMUS/SCI/17;
the Danish Ministry of Science;
the Estonian Ministry of Education and Research through grant IUT40-1;
the European Commission's Sixth Framework Programme through the European Leadership in Space Astrometry (ELSA) Marie Curie Research Training Network (MRTN-CT-2006-033481), through Marie Curie project PIOF-GA-2009-255267 (Space AsteroSeismology \& RR Lyrae stars, SAS-RRL), and through a Marie Curie Transfer-of-Knowledge (ToK) fellowship (MTKD-CT-2004-014188); the European Commission's Seventh Framework Programme through grant FP7-606740 (FP7-SPACE-2013-1) for the Gaia European Network for Improved data User Services (GENIUS) and through grant 264895 for the Gaia Research for European Astronomy Training (GREAT-ITN) network;
the European Research Council (ERC) through grants 320360 and 647208 and through the European Union's Horizon 2020 research and innovation programme through grants 670519 (Mixing and Angular Momentum tranSport of massIvE stars - MAMSIE) and 687378 (Small Bodies: Near and Far);
the European Science Foundation (ESF), in the framework of the Gaia Research for European Astronomy Training Research Network Programme (GREAT-ESF);
the European Space Agency (ESA) in the framework of the Gaia project, through the Plan for European Cooperating States (PECS) programme through grants for Slovenia, through contracts C98090 and 4000106398/12/NL/KML for Hungary, and through contract 4000115263/15/NL/IB for Germany;
the European Union (EU) through a European Regional Development Fund (ERDF) for Galicia, Spain;
the Academy of Finland and the Magnus Ehrnrooth Foundation;
the French Centre National de la Recherche Scientifique (CNRS) through action 'D\'efi MASTODONS' the Centre National d'Etudes Spatiales (CNES), the L'Agence Nationale de la Recherche (ANR)  'Investissements d'avenir' Initiatives D'EXcellence (IDEX) programme Paris Sciences et Lettres (PSL**) through grant ANR-10-IDEX-0001-02, the ANR D'efi de tous les savoirs' (DS10) programme through grant ANR-15-CE31-0007 for project 'Modelling the Milky Way in the Gaia era' (MOD4Gaia), the R\'egion Aquitaine, the Universit\'e de Bordeaux, and the Utinam Institute of the Universit\'e de Franche-Comt\'e, supported by the R\'egion de Franche-Comt\'e and the Institut des Sciences de l'Univers (INSU);
the German Aerospace Agency (Deutsches Zentrum f\"ur Luft- und Raumfahrt e.V., DLR) through grants 50QG0501, 50QG0601, 50QG0602, 50QG0701, 50QG0901, 50QG1001, 50QG1101, 50QG1401, 50QG1402, 50QG1403, and 50QG1404 and the Centre for Information Services and High Performance Computing (ZIH) at the Technische Universit\"at (TU) Dresden for generous allocations of computer time;
the Hungarian Academy of Sciences through the Lend\"ulet Programme LP2014-17 and the J\'anos Bolyai Research Scholarship (L. Moln\'ar and E. Plachy) and the Hungarian National Research, Development, and Innovation Office through grants NKFIH K-115709, PD-116175, and PD-121203;
the Science Foundation Ireland (SFI) through a Royal Society - SFI University Research Fellowship (M. Fraser);
the Israel Science Foundation (ISF) through grant 848/16;
the Agenzia Spaziale Italiana (ASI) through contracts I/037/08/0, I/058/10/0, 2014-025-R.0, and 2014-025-R.1.2015 to the Italian Istituto Nazionale di Astrofisica (INAF), contract 2014-049-R.0/1/2 to INAF dedicated to the Space Science Data Centre (SSDC, formerly known as the ASI Sciece Data Centre, ASDC), and contracts I/008/10/0, 2013/030/I.0, 2013-030-I.0.1-2015, and 2016-17-I.0 to the Aerospace Logistics Technology Engineering Company (ALTEC S.p.A.), and INAF;
the Netherlands Organisation for Scientific Research (NWO) through grant NWO-M-614.061.414 and through a VICI grant (A. Helmi) and the Netherlands Research School for Astronomy (NOVA);
the Polish National Science Centre through HARMONIA grant 2015/18/M/ST9/00544 and ETIUDA grants 2016/20/S/ST9/00162 and 2016/20/T/ST9/00170;
the Portugese Funda\c{c}\~ao para a Ci\^encia e a Tecnologia (FCT) through grant SFRH/BPD/74697/2010; the Strategic Programmes UID/FIS/00099/2013 for CENTRA and UID/EEA/00066/2013 for UNINOVA;
the Slovenian Research Agency through grant P1-0188;
the Spanish Ministry of Economy (MINECO/FEDER, UE) through grants ESP2014-55996-C2-1-R, ESP2014-55996-C2-2-R, ESP2016-80079-C2-1-R, and ESP2016-80079-C2-2-R, the Spanish Ministerio de Econom\'\i a, Industria y Competitividad through grant AyA2014-55216, the Spanish Ministerio de Educaci\'on, Cultura y Deporte (MECD) through grant FPU16/03827, the Institute of Cosmos Sciences University of Barcelona (ICCUB, Unidad de Excelencia 'Mar\'\i a de Maeztu') through grant MDM-2014-0369, the Xunta de Galicia and the Centros Singulares de Investigaci\'on de Galicia for the period 2016-2019 through the Centro de Investigaci\'on en Tecnolog\'{\i}as de la Informaci\'on y las Comunicaciones (CITIC), the Red Espa\~nola de Supercomputaci\'on (RES) computer resources at MareNostrum, and the Barcelona Supercomputing Centre - Centro Nacional de Supercomputaci\'on (BSC-CNS) through activities AECT-2016-1-0006, AECT-2016-2-0013, AECT-2016-3-0011, and AECT-2017-1-0020;
the Swedish National Space Board (SNSB/Rymdstyrelsen);
the Swiss State Secretariat for Education, Research, and Innovation through the ESA PRODEX programme, the Mesures d'Accompagnement, the Swiss Activit\'es Nationales Compl\'ementaires, and the Swiss National Science Foundation;
the United Kingdom Rutherford Appleton Laboratory, the United Kingdom Science and Technology Facilities Council (STFC) through grant ST/L006553/1, the United Kingdom Space Agency (UKSA) through grant ST/N000641/1 and ST/N001117/1, as well as a Particle Physics and Astronomy Research Council Grant PP/C503703/1.



%\bibliographystyle{mnras}


%%%%%%%%%%%%%%%%%%%%%%%%%%%%%%%%%%%%%%%%%%%%%%%%%%

%%%%%%%%%%%%%%%%% APPENDICES %%%%%%%%%%%%%%%%%%%%%
%\clearpage

% Don't change these lines
\bsp	% typesetting comment
\label{lastpage}
\end{document}